\documentclass[12pt]{article}
\pdfoutput=1
\usepackage{color}
\usepackage[dvipsnames]{xcolor}
\usepackage[linktoc=page,bookmarks=false,colorlinks=false,
linkbordercolor=RoyalBlue,citebordercolor=ForestGreen,urlbordercolor=CornflowerBlue]{hyperref}
\usepackage{amssymb,amsmath,bm,bbm}
\usepackage{epsf}
\usepackage{epsfig}
\usepackage{afterpage}
\usepackage{longtable}
\usepackage[compress]{cite}
\usepackage[normalem]{ulem}
\usepackage{latexsym,mathrsfs,dsfont}
\usepackage{graphics}
\usepackage{url}
\usepackage{paralist}
\usepackage{bbold}
\usepackage{tocloft}

\newcommand{\ord}{\mathcal{O}}

\newcommand{\IM}{\rm{Im}}

\newcommand{\tev}{\, {\rm TeV}}
\newcommand{\gev}{\, {\rm GeV}}
\newcommand{\mev}{\, {\rm MeV}}

\newcommand{\vcb}{|V_{cb}|}

\newcommand{\vub}{|V_{ub}|}

\newcommand{\vus}{|V_{us}|}

\newcommand{\bsi}{B_6^{(1/2)}}
\newcommand{\bei}{B_8^{(3/2)}}

\def\epe{\varepsilon'/\varepsilon}
\newcommand{\beq}{\begin{equation}}
\newcommand{\eeq}{\end{equation}}
\newcommand{\be}{\begin{equation}}
\newcommand{\ee}{\end{equation}}
\newcommand{\bi}{\begin{itemize}}
\newcommand{\ei}{\end{itemize}}
\newcommand{\ba}{\begin{array}}
\newcommand{\ea}{\end{array}}
\newcommand{\beqa}{\begin{eqnarray}}
\newcommand{\eeqa}{\end{eqnarray}}
\newcommand{\bea}{\begin{eqnarray}}
\newcommand{\eea}{\end{eqnarray}}
\newcommand{\beqn}{\begin{eqnarray}}
\newcommand{\eeqn}{\end{eqnarray}}

\newcommand{\D}{\Delta}

\newcommand{\eps}{\epsilon}

\definecolor{red}{cmyk}{0,1,1,0.4}

\def\kpn{K^+\rightarrow\pi^+\nu\bar\nu}
\def\klpn{K_{L}\rightarrow\pi^0\nu\bar\nu}
\def\B{\mathcal{B}}

\def\D{\bf\color{blue}}

\usepackage{fancyhdr}
\advance \headheight by 3.0truept       

\begin{document}

\begin{flushright}
    {FLAVOUR(267104)-ERC-105}
\end{flushright}

\medskip

\begin{center}
{\Large\bf
\boldmath{$K\to\pi\nu\bar\nu$  and $\epe$ in Simplified New Physics Models}}
\\[0.8 cm]
{\bf Andrzej~J.~Buras, Dario Buttazzo and Robert Knegjens
 \\[0.5 cm]}
{\small
TUM Institute for Advanced Study, Lichtenbergstr. 2a, D-85748 Garching, Germany\\
Physik Department, Technische Universit\"at M\"unchen,
James-Franck-Stra{\ss}e, \\D-85748 Garching, Germany}
\end{center}

\vskip0.41cm

\abstract{%
\noindent
The decays $\kpn$ and $\klpn$, being  the theoretically cleanest rare decays 
of mesons, are very sensitive probes of New Physics (NP). In view of 
the excellent prospects of reaching the Standard Model (SM) sensitivity for 
$\kpn$ by the NA62 experiment at CERN and for $\klpn$ by the KOTO experiment at 
 J-PARC, we study them in the simplest extensions of the SM in which 
 stringent correlations between these two decays and other flavour observables 
are present. We first consider simple models with tree-level $Z$ 
and $Z^\prime$ contributions in which either Minimal Flavour Violation (MFV) or a $U(2)^3$ symmetry is imposed on 
the quark flavour-violating couplings. We then compare the resulting correlations with 
those present in generic models in which the latter couplings are arbitrary, subject 
to the constraints from $\Delta F=2$ processes, electroweak and collider 
data. Of particular interest are the correlations with $\epe$ and $K_L\to \mu^+\mu^-$ which limit the size of NP contributions 
to $\kpn$ and $\klpn$, depending on the Dirac structure of couplings and 
the relevant operators. But in MFV models also the constraint from 
$B_s\to\mu^+\mu^-$ turns out to be important. We take into account the 
recent results from lattice QCD and large $N$ approach that indicate $\epe$ 
in the SM to be significantly below the data.
While in many models the enhancement of $\epe$ implies the suppression of $\klpn$, we present two models in which 
 $\epe$  and $\klpn$ can be simultaneously enhanced relative to 
SM predictions. A correlation between $\kpn$ and  
$B\to K(K^*)\mu^+\mu^-$, found by us in the simple models considered here, should be of interest 
for NA62 and LHCb experimentalists at CERN in the coming years. The one 
with $B\to K(K^*)\nu\bar\nu$ will be tested at Belle II.}

\vfill

\hrule

\medskip

{\noindent\hspace{-18pt}\footnotesize E-mail: \href{mailto:andrzej.buras@tum.de}{andrzej.buras@tum.de}, \href{mailto:dario.buttazzo@tum.de}{dario.buttazzo@tum.de}, \href{mailto:robert.knegjens@tum.de}{robert.knegjens@tum.de}}

\thispagestyle{empty}
\newpage
\setcounter{page}{1}

\renewcommand{\cftsecfont}{\bfseries\boldmath}
\tableofcontents


\section{Introduction}
After more than twenty years of waiting \cite{Buchalla:1993bv}, the prospects of measuring the 
branching ratios for two {\it golden} modes
$\kpn$ and $\klpn$ within this decade are very good. Indeed,
the NA62 experiment at CERN is expected to measure the $\kpn$ branching ratio with the precision of $\pm10\%$ \cite{Rinella:2014wfa,Romano:2014xda}, and the KOTO experiment at J-PARC should make a significant progress in measuring the branching ratio for $\klpn$ \cite{Komatsubara:2012pn,Shiomi:2014sfa}.

These decays are theoretically 
very clean and their branching ratios have been calculated within  the SM including NNLO QCD corrections \cite{Gorbahn:2004my,Buras:2005gr,Buras:2006gb} and  NLO electroweak corrections \cite{Brod:2008ss,Brod:2010hi,Buchalla:1997kz}.
Moreover, extensive calculations of isospin breaking effects and 
non-perturbative effects have been done \cite{Isidori:2005xm,Mescia:2007kn}. 
Therefore, once the CKM parameters $\vcb$, $\vub$ and $\gamma$  will be 
precisely determined in tree-level decays, these two decays will 
offer an excellent probe of the physics beyond the SM.
Reviews of these two decays can be found in 
\cite{Buras:2004uu,Komatsubara:2012pn,Buras:2013ooa,Blanke:2013goa,Smith:2014mla}. 

In a recent paper \cite{Buras:2015qea} we have reviewed the status of these decays within the SM 
taking into account all presently available information from other observables 
and lattice QCD. In calculating the branching ratios for these decays we followed two strategies:

{\bf Strategy A:} in which the  CKM matrix is determined using tree-level measurements of 
\be\label{STRA}
\vus,\qquad \vcb, \qquad \vub, \qquad \gamma,
\ee
where $\gamma$ is one of the angles of the unitarity triangle. As new physics (NP) seems to be by now well separated from the electroweak scale, this determination is  likely not polluted by NP contributions allowing the determination of 
{\it true} values of all elements of the CKM matrix. Inserting these values 
into the known expressions for the relevant branching ratios allowed us to determine the SM values for these branching ratios independently of whether NP is present at short distance scales or not. We found
\begin{align}\label{PREDA}
    \mathcal{B}(\kpn) &= \left(8.4 \pm 1.0\right) \times 10^{-11}, \\
    \mathcal{B}(\klpn) &= \left(3.4\pm 0.6\right) \times 10^{-11}.    
\end{align}
This strategy is 
clearly optimal as it allows to predict {\it true} SM values of these branching 
ratios. 

{\bf Strategy B:} in which it is assumed that the SM is the whole story and 
the values of CKM parameters are extracted from $\Delta F=2$ observables, in 
particular $\varepsilon_K$, $\Delta M_s$, $\Delta M_d$ and mixing induced CP 
asymmetries $S_{\psi K_S}$ and $S_{\psi\phi}$. Having more constraints, 
 more accurate values of $\vcb$, $\vub$ and $\gamma$ than in strategy A 
could be found implying significantly more accurate predictions
\begin{align}\label{PREDB}
    \mathcal{B}(\kpn) &= \left(9.1 \pm 0.7\right) \times 10^{-11}, \\
    \mathcal{B}(\klpn) &= \left(3.0\pm 0.3\right) \times 10^{-11}.    
\end{align}
These latter  results are useful in the sense that in the case of future measurements 
of these two branching ratios differing from them would signal the presence of 
NP but this NP would not necessarily be contributing to these two decays as 
it could also pollute the determination of CKM parameters through loop decays.

Evidently, strategy A is superior to strategy B  in the context of NP analyses,  since it allows to determine the CKM matrix elements independently of NP effects which may depend on a large number of parameters.  But in a given NP model, in which contributions to rare processes involve 
only a small number of new parameters in addition to the SM ones, strategy B could also be 
efficiently used.  However, in the present paper we will exclusively use 
the strategy A.

The decays  $\kpn$ and $\klpn$ have been studied over many years in various 
concrete extensions of the SM. A review of the analyses performed until 
August 2007 can be found in \cite{Buras:2004uu}. More recent reviews can be 
found in \cite{Buras:2010wr,Buras:2012ts,Buras:2013ooa,Blanke:2013goa,Smith:2014mla}. Most extensive analyses have been performed in supersymmetric models \cite{Buras:1997ij,Colangelo:1998pm,Buras:1999da,Buras:2004qb,Crivellin:2011sj}, 
the Littlest Higgs (LH) model without T-parity \cite{Buras:2006wk}, the LH model with T-parity (LHT) \cite{Blanke:2009am,Blanke:2015wba}, Randall-Sundrum models \cite{Blanke:2008yr,Bauer:2009cf}, models with partial compositeness \cite{Straub:2013zca} and  331 models \cite{Buras:2012dp,Buras:2013dea,Buras:2014yna}. All these models 
contain several new parameters related to couplings and masses of new fermions,
 gauge bosons and scalars and the analysis of  $\kpn$ and $\klpn$  requires 
the inclusion of all constraints on couplings and masses of these particles and 
consequently is rather involved. Moreover, the larger number of parameters 
present in these models does not presently allow for clear cut conclusions beyond rough bounds on the size of NP contributions to  $\kpn$ and $\klpn$. 

Therefore, we think that presently in order to get a better insight into the structure of the possible impact of NP on  $\kpn$ and $\klpn$ decays, and in particular on the correlation between them and other observables, it is useful to consider models with a only small number of parameters. 
With this idea in mind we will consider:
\begin{itemize}
\item
General classes of models based on a $U(3)^3$ flavour symmetry (MFV), illustrating 
them by means of two specific models in which quark flavour violating couplings 
of  $Z$  and of a heavy $Z^\prime$  are consistent with this symmetry.
\item
Models in which the flavour symmetry $U(3)^3$ is reduced to $U(2)^3$, illustrating the results again by means of two simple $Z$ and $Z^\prime$ models.
\item
The $Z$ and $Z^\prime$ models with tree-level FCNCs in which the quark 
couplings are arbitrary 
subject to available constraints from other decays. In particular in this 
case we will include right-handed currents which are absent in MFV and strongly 
suppressed in the simplest $U(2)^3$ models.
\end{itemize}
Note that in each case we consider as benchmarks $Z$ and $Z^\prime$ models with tree-level FCNCs to quarks, and flavour-conserving, as well as flavour universal, couplings to leptons.
Neglecting the tiny neutrino masses, one can assume NP to have only left-handed vector couplings to the neutrino pair, and ignore scalar currents. Therefore simplified models involving gauge-bosons form a good generalisation of the more specific NP models available.
The simplified $Z$ can mimic modified $Z$ penguins for instance, occurring in supersymmetric models for example, while a $Z'$-like particle occurs in several of the other models listed earlier.

In addition to  $\kpn$ and $\klpn$ the ratio $\epe$ belongs to the most
prominent observables in $K$-meson physics. It is also very sensitive to NP contributions, but is unfortunately subject to large hadronic uncertainties present in 
the matrix elements of QCD and electroweak penguin operators. Moreover, 
strong cancellations between these two contributions make precise predictions 
for $\epe$ in the SM and its various extensions difficult. Reviews of $\epe$ can be found in \cite{Bertolini:1998vd,Buras:2003zz,Pich:2004ee,Cirigliano:2011ny,Bertolini:2012pu}. The most recent analyses of $\epe$ within $Z(Z^\prime)$ and 331 models have been presented in \cite{Buras:2014sba} and \cite{Buras:2014yna}, respectively. See also our SM analysis in \cite{Buras:2015qea}.

Most importantly, improved anatomy of $\epe$ within the SM 
have been presented in \cite{Buras:2015yba}. It was triggered by the first result on $\epe$ from the RCB-UKQCD lattice collaboration \cite{Bai:2015nea}, which indicated that $\epe$ in the SM could be significantly below the data, but the large 
 theoretical uncertainties in this calculation did not yet allow for firm 
conclusions. These uncertainties have been significantly reduced in 
\cite{Buras:2015yba} through the extraction of a number of hadronic matrix elements of
contributing operators from the CP-conserving $K\to\pi\pi$ data. Parallel 
to this study an important upper bound for the contribution of QCD penguins 
to $\epe$ has been derived from the large N approach \cite{Buras:2015xba}. 
The analysis in \cite{Buras:2015yba} combined with the bound in  \cite{Buras:2015xba} demonstrates  
that indeed $\epe$ in the SM could turn out to be significantly lower than its experimental value. We will be more explicit about this in section~\ref{sec:3a}.

Now, in most extensions of the SM found in the literature the enhancement of $\epe$ through NP usually implies the suppression of the branching ratio for $\klpn$.  But, as we will demonstrate in Section~\ref{sec:lat} simplified models 
can be constructed in which 
$\epe$ and the branching ratio for $\klpn$ can be simultaneously 
enhanced over their SM values.

Our paper is organised as follows. 
In section~\ref{sec:2} we collect basic formulae for $\kpn$ and $\klpn$ valid in any extension of the SM and discuss their general properties. 
In section~\ref{sec:3} we formulate the simple $Z$ and $Z^\prime$ models in 
question. 
In section~\ref{sec:3a} we recall some aspects of $\epe$ concentrating on the
simplified models of the previous section. In particular  we present 
two simplified models in which $\epe$,  $\mathcal{B}(\kpn)$ and $\mathcal{B}(\klpn)$ can be enhanced simultaneously over their SM values.  
In section~\ref{sec:4} we present formulae  for various decays and observables in the simplified models of section~\ref{sec:3} and discuss their correlations with $\kpn$ and $\klpn$. This includes $b\to s\ell^+\ell^-$ transitions,  $B\to K(K^*)\nu\bar\nu$  and $K_L\to\mu^+\mu^-$.  $K_L\to\mu^+\mu^-$ 
plays an important role in constraining the allowed values of $\mathcal{B}(\kpn)$. 
 While some numerical results will be shown already in previous sections 
the main numerical analysis of the models of section~\ref{sec:3}  
is presented in section~\ref{sec:6}.  
We conclude in section~\ref{sec:8}.


\boldmath
\section{General Formulae and Properties}\label{sec:2}
\unboldmath

\subsection{General Expressions}
The branching ratios for $\kpn$  and $\klpn$ in any extension of the SM in 
which light neutrinos couple only to left-handed currents  are given as 
follows
\begin{align}
\mathcal{B}(K^+\to\pi^+\nu\bar\nu)&=\kappa_+ (1+\Delta_\text{EM})\cdot
\left[\left(\frac{{\rm Im}\,X_{\rm eff}}{\lambda^5}\right)^2+
\left(\frac{{\rm Re}\,\lambda_c}{\lambda}P_c(X)+
\frac{{\rm Re}\,X_{\rm eff}}{\lambda^5}\right)^2\right]\,,\label{bkpnn}\\
\mathcal{B}(\klpn)&=\kappa_L\cdot
\left(\frac{{\rm Im}\,X_{\rm eff}}{\lambda^5}\right)^2\,,\label{bklpn}
\end{align}
where \cite{Mescia:2007kn}
\begin{align}
\kappa_+ &= (5.173\pm 0.025 )\cdot 10^{-11}\left[\frac{\lambda}{0.225}\right]^8, & \Delta_\text{EM} &=-0.003\,,\label{kapp}\\
\kappa_L &= (2.231\pm 0.013)\cdot 10^{-10}\left[\frac{\lambda}{0.225}\right]^8. & \label{kapl}
\end{align}
and $\lambda_i=V^*_{is}V_{id}$ are the CKM factors.
For the charm contribution, represented by $P_c(X)$, the calculations in 
\cite{Buras:2005gr,Buras:2006gb,Brod:2008ss,Isidori:2005xm,Mescia:2007kn}
imply \cite{Buras:2015qea}
\be\label{PCFINAL}
P_c(X)= 0.404\pm 0.024,
\ee
where the error is dominated by the  long distance uncertainty estimated in \cite{Isidori:2005xm}. In what follows we will assume that NP does not modify this value, which turns 
out to be true in all extensions of the SM we know about. Such contributions can be in any case  absorbed into the function $X_{\rm eff}$. The latter function 
that describes pure  short distance contributions from top quark exchanges and 
NP is given by 
\be\label{XK}
X_{\rm eff} = V_{ts}^* V_{td} (X_{L}(K) + X_{R}(K))\equiv 
V_{ts}^* V_{td} X_L^{\rm SM}(K) ( 1 +\xi e^{i\theta}).
\ee
The functions $X_{L}(K)$ and $X_{R}(K)$ summarise the contributions from 
left-handed and right-handed quark currents, respectively. In the SM only 
$X_{L}(K)$ is non-vanishing and is given by \cite{Buras:2015qea}
\be\label{XSM}
X_L^{\rm SM}(K)= 1.481\pm 0.005_\text{th}\pm 0.008_\text{exp}=1.481 \pm 0.009.
\ee

One can also express the function $X_{\rm eff}$ as a function of the branching ratios $\B(\kpn)$ and $\B(\klpn)$, which is useful for the study of correlations of the latter with other flavour observables. One has, directly from \eqref{bkpnn}, \eqref{bklpn},
\begin{align}
{\rm Re}\,X_{\rm eff} &= -\lambda^5\left[\frac{\B(\kpn)}{\kappa_+(1 + \Delta_{\rm EM})} - \frac{\B(\klpn)}{\kappa_L}\right]^{1/2} - \lambda^4{\rm Re}\,\lambda_c P_c(X)\,,\label{ReX}\\
{\rm Im}\,X_{\rm eff} &= \lambda^5\left[\frac{\B(\klpn)}{\kappa_L}\right]^{1/2}\,.\label{ImX}
\end{align}
In choosing the signs in these formulae we assumed that NP contributions 
 do not reverse the sign of SM functions. For more general expressions 
admitting such a possibility see  \cite{Buras:2001af}. At the Grossmann-Nir 
bound \cite{Grossman:1997sk} the square root in (\ref{ReX}) vanishes.


\subsection{Basic Properties}

\begin{figure}[t]
\centering%
\includegraphics[width=0.6\textwidth]{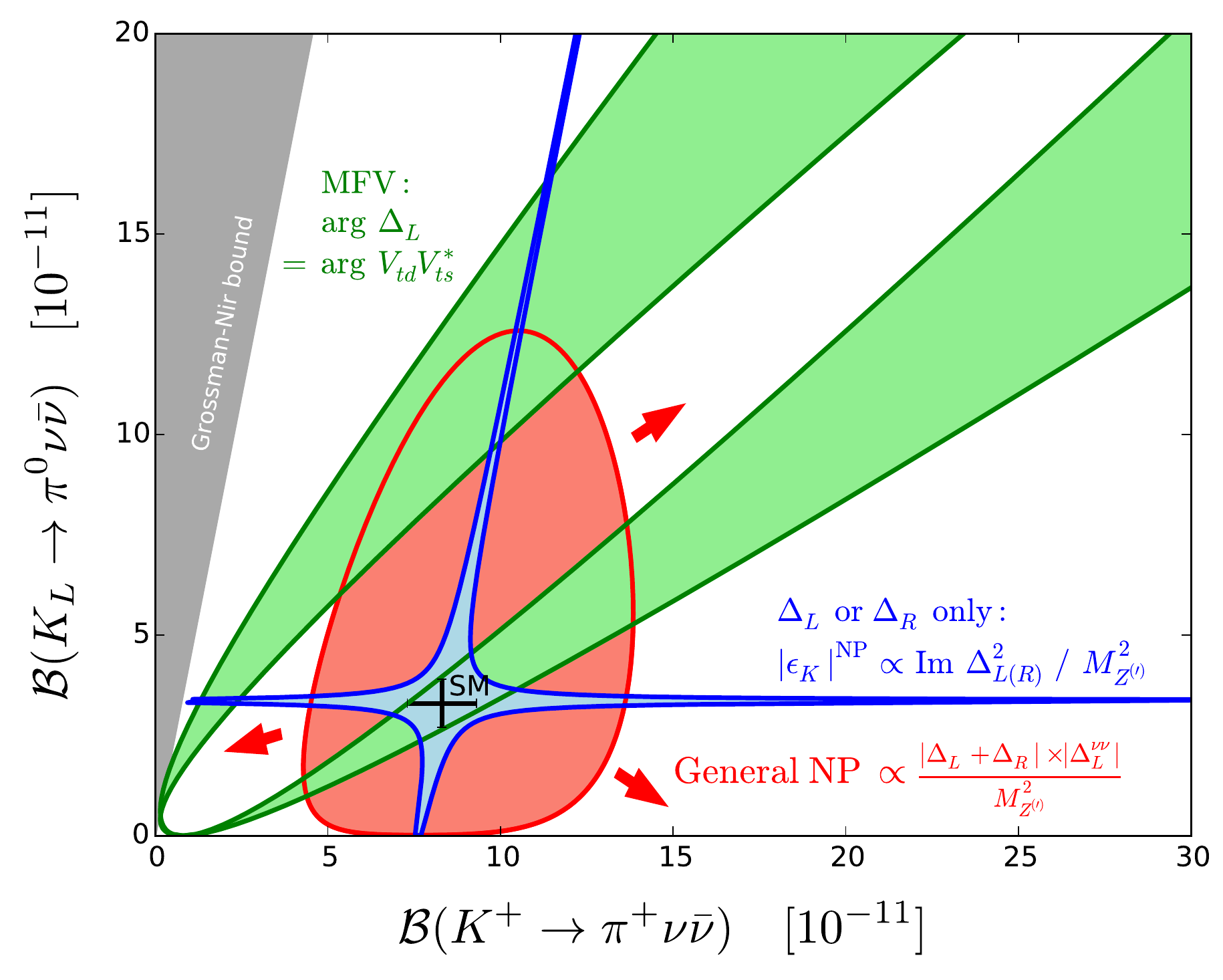}%
\caption{\it Illustrations of common correlations in the $\mathcal{B}(\kpn)$ versus $\mathcal{B}(\klpn)$ plane. The expanding red region illustrates the lack of correlation for models with general LH and RH NP couplings. The green region shows the correlation present in models obeying CMFV. The blue region shows the correlation induced by the constraint from $\varepsilon_K$ if only LH or RH couplings  are present. \label{fig:illustrateEpsK}}
\end{figure}

The correlation between $\mathcal{B}(\kpn)$ and $\mathcal{B}(\klpn)$ 
depends on the short distance dynamics, encapsulated in the two real parameters $\xi$ and $\theta$ that vanish in the SM. Measuring these branching ratios one day will allow to determine those parameters and, comparing them with their expectations in concrete models, obtain insight into the flavour structure of the NP contributions. Those can be dominated by left-handed currents, by right-handed currents, or by both with similar magnitudes and phases. In general one can distinguish between three classes of models \cite{Blanke:2009pq}:

\begin{enumerate}
\item
Models with a CKM-like structure of flavour interactions. If based on flavour symmetries only, they include MFV and $U(2)^3$ models \cite{Barbieri:2014tja}.
In this case the function $X_L(K)$ is real and $X_R(K)=0$. 
There is then only one variable to our disposal, the value of  $X_L(K)$,  and the only allowed values of both branching ratios are on the  green  branches in figure~\ref{fig:illustrateEpsK}. 
But due to stringent correlations with other observables present in this class of models, only certain ranges for $\mathcal{B}(\kpn)$ and $\mathcal{B}(\klpn)$
are still allowed, which we will determine in the context of our analysis.
\item
Models with new flavour and CP-violating interactions in which either
left-handed currents or right-handed currents fully dominate, implying that 
left-right operator contributions to $\varepsilon_K$ can be neglected. In 
this case there is a strong correlation between NP contributions to $\varepsilon_K$ and $K\to\pi\nu\bar\nu$ and the $\varepsilon_K$ constraint implies 
the blue branch structure shown in figure~\ref{fig:illustrateEpsK}.  
On the horizontal branch NP contribution to $K\to\pi\nu\bar\nu$ is real and therefore vanishes in the case of $\klpn$. On the second branch  NP
contribution is purely imaginary and this branch is parallel to the Grossman-Nir (GN) bound \cite{Grossman:1997sk}. In practice, due to uncertainties in $\varepsilon_K$, there are moderate deviations from this structure which is characteristic for the LHT model \cite{Blanke:2009am}, or $Z$ or $Z^\prime$ FCNC scenarios with either pure LH or RH couplings \cite{Buras:2012jb,Buras:2014zga}.
\item
If left-right operators have significant contribution to $\varepsilon_K$ or 
generally if the correlation between $\varepsilon_K$ and $K\to\pi\nu\bar\nu$ 
is weak or absent, the two branch structure is also absent. Dependent on 
the values of $\xi$ or $\theta$, any value of $\mathcal{B}(\kpn)$ and  $\mathcal{B}(\klpn)$ is in principle possible. The red region in figure~\ref{fig:illustrateEpsK} shows the resulting structure for a fixed value of $\xi$ and $0\le\theta\le 2\pi$. Randall-Sundrum models with 
custodial protection (RSc) belong to this class of models \cite{Blanke:2008yr}. 
However, it should be kept in mind that usually the removal of the correlation with $\varepsilon_K$ requires subtle cancellations between different 
contributions to $\varepsilon_K$ and consequently some tuning of parameters \cite{Blanke:2008yr,Buras:2014zga}.
\end{enumerate}

Unfortunately, on the basis of only these two 
branching ratios alone, it is not possible to find out how important the 
contributions of right-handed currents are, as their effects are hidden in 
a single function $X_{\rm eff}$. In this sense the decays governed 
by $b\to s \nu\bar\nu$ transitions, which will also enter our analysis, 
are complementary, and the correlation between $K\to\pi\nu\bar\nu$ decays and 
$B\to K(K^*)\nu\bar\nu$, as well as $B_{s,d}\to\mu^+\mu^-$, can help in identifying 
the presence or absence of right-handed currents.


\section{Simplified Models}\label{sec:3}

In studying correlations between various decays it is important to remember that
\begin{itemize}
\item
Correlations between decays of different mesons test the flavour structure 
of couplings or generally flavour symmetries.
\item
Correlations between decays of a given meson test the Dirac structure of 
couplings.
\end{itemize}
We will look at the first correlations by comparing those within 
MFV models based on a $U(3)^3$ flavour symmetry with the ones present in 
models with a minimally broken $U(2)^3$ flavour symmetry \cite{Barbieri:2011ci,Barbieri:2012uh}.
In the latter case we will work at leading order in the breaking of the symmetry, and therefore assume that only the left-handed quark couplings are relevant, as in MFV. We will then extend the analysis to more general models with generic flavour structure.


\boldmath
\subsection{$Z$ models with flavour symmetries}
\unboldmath

In order to exhibit correlations of $\kpn$ and $\klpn$ decays with other observables we will first
consider two simple $Z$ models in which the quark flavour violating 
couplings are consistent either with a $U(3)^3$ or with a $U(2)^3$ symmetry. 
These models are very restrictive as the $Z$ mass and its couplings to leptons 
are known. In particular, in the conventions of \cite{Buras:2012jb} for the couplings $\Delta(Z)$ of the $Z$ boson to fermions,
\be\label{LAZ}
\Delta_L^{\nu\bar\nu}(Z)=\Delta_A^{\mu\bar\mu}(Z)=\frac{g}{2c_W}=0.372\,.
\ee
However, in order to be able to generalise our analysis straightforwardly 
to the $Z^\prime$ case, we will use the general expressions for these lepton
couplings.

We will then find that  in the case of MFV there is only one new real parameter $a$   and in 
the $U(2)^3$ case there are three new real parameters: real $a$ and a complex 
$b$. 


\boldmath
\subsubsection{$U(3)^3$ case}
\unboldmath

In this case the $Z$ quark flavour violating couplings are given respectively for the three meson systems $(K,B_d,B_s)$ as follows:
\begin{align}\label{MFVcouplings}
\Delta_L^{sd}(Z) &= a V_{ts}^*V_{td}, & \Delta_L^{db}(Z) &= a V_{td}^*V_{tb}, & \Delta_L^{sb}(Z) &= a V_{ts}^*V_{tb},
\end{align}
where $a$ is flavour-universal and real.

The presence of tree-level $Z$ contributions in various flavour observables 
can be summarised by shifts in the master functions $S$, $X$ and $Y$ which 
enter respectively the expressions for quark mixing ($\Delta F=2$) and branching ratios 
for meson decays with $\nu\bar\nu$ and $\mu^+\mu^-$ in the final state. 

The couplings in \eqref{MFVcouplings} imply then:
\begin{equation}\label{DeltaS}
\Delta S(K) = \Delta S(B_d)=\Delta S(B_s) \equiv \Delta S =a^2 \frac{4\tilde r}{M_{Z}^2g_{\text{SM}}^2}
\end{equation}
where
\be\label{gsm}
g_{\text{SM}}^2=
4 \frac{M_W^2 G_F^2}{2 \pi^2} = 1.78137\times 10^{-7} \gev^{-2}\,,
\ee
with $G_{F}$ being the Fermi constant.  $\tilde r= 1.068$ is a QCD 
correction \cite{Buras:2012jb}.

Similarly, 
\be\label{XSHIFT}
\Delta X_L(K)=\Delta X_L(B_d)=\Delta X_L(B_s)\equiv \Delta X=a \frac{\Delta_L^{\nu\bar\nu}(Z)}{M_{Z}^2g_{\text{SM}}^2},
\ee
and 
\be
\Delta Y_A(K)=\Delta Y_A(B_d)=\Delta Y_A(B_s)\equiv \Delta Y =a \frac{\Delta_A^{\mu\bar\mu}(Z)}{M_{Z}^2g_{\text{SM}}^2}\,.
\ee
We observe very strong correlations between the three meson systems. This model 
has only one new real parameter $a$  with respect to the SM, which could be positive or negative. In fact, using the equality of the $Z$ couplings in (\ref{LAZ})
and eliminating the parameter $a$ we find a very stringent relation
\be\label{RR1}
\Delta X=\Delta Y= \pm 4.67 \sqrt{\Delta S},
\ee
where the sign corresponds to two possible signs of $a$. The consequences of 
this relation are rather profound. In particular:
\begin{itemize}
\item
The size of possible effects in rare decays is strongly bounded by the allowed
universal shift in the box function $S$.
\item
However, as $S_\text{SM}>X_\text{SM}>Y_\text{SM}>0$, NP generically affects, in this scenario, rare 
decays stronger than  particle-antiparticle mixing.
\item
While the flavour universal shifts $\Delta X$ and $\Delta Y$ can have generally both signs, 
with the real parameter $a$, the universal shifts $\Delta S$ are strictly positive in agreement with the general discussion in \cite{Blanke:2006yh}. This means 
that $\Delta M_{s,d}$ and $\varepsilon_K$ can only be enhanced in this scenario, and this happens in a correlated manner.
\item
Due to the present data on $B_s\to\mu^+\mu^-$ the shift $\Delta Y< 0$ is favoured, implying {\it suppression} of all rare decay branching ratios governed 
by the functions $X$ and $Y$. Moreover, the amounts of these suppressions are 
correlated with each other. We stress that this property is characteristic for
tree-level $Z$ exchange and originates in the signs of the leptonic couplings 
in (\ref{LAZ}).
\item
 As in the SM $X_\text{SM}>Y_\text{SM}$, NP affects stronger  decays 
 with $\mu\bar\mu$ in the final state than  those with $\nu\bar\nu$.
\end{itemize}
Our numerical analysis in section~\ref{sec:6} will show that in this scenario 
NP effects are generally below $50\%$ at the level of the branching ratios.


\boldmath
\subsubsection{$U(2)^3$ case}
\unboldmath

The $Z$ couplings in (\ref{MFVcouplings}) are now modified to 
\be\label{U23couplings}
\Delta_L^{sd}(Z)=a V_{ts}^*V_{td}, \qquad \Delta_L^{db}(Z)= b V_{td}^*V_{tb}, \qquad \Delta_L^{sb}(Z)= b V_{ts}^*V_{tb}\,,
\ee
with $b\not= a$ being a complex number. Therefore, compared with the $U(3)^3$ case, $b$ represents two 
new real parameters: its absolute value, and the phase which has impact on 
CP violation in $B_{s,d}$ systems.
In this case the correlation between the $K$ system and the $B_{s,d}$ systems is broken.
For the $K$ system the MFV formulae remain unchanged, while now
\begin{align}
\Delta S(B_d)&=\Delta S(B_s)\equiv \Delta S(B)=(b^*)^2 \frac{4\tilde r}{M_{Z}^2g_{\text{SM}}^2},\\
\Delta X_L(B_d)&=\Delta X_L(B_s) \equiv \Delta X(B)=b \frac{\Delta_L^{\nu\bar\nu}(Z)}{M_{Z}^2g_{\text{SM}}^2},\\
\Delta Y_A(B_d)&=\Delta Y_A(B_s)\equiv \Delta Y(B)=b \frac{\Delta_A^{\mu\bar\mu}(Z)}{M_{Z}^2g_{\text{SM}}^2}\,.
\end{align}
We find then
\be\label{RR2}
\Delta X(B)=\Delta Y(B)= \pm 4.67 \sqrt{\Delta S(B)^*}.
\ee
Moreover, writing the total $S$ function as
\be
S(B)=S_{\rm SM}+\Delta S(B)=|S(B)|e^{-i2\varphi}\,,
\ee
where a non-zero $\varphi$ is generated by quark flavour violating $Z$ couplings, we find the known 
anti-correlation between mixing induced  CP asymmetries in $B_d$ and $B_s$ systems respectively:
\be\label{CPASYM}
S_{\psi K_S}=\sin(2\beta+2\varphi),\qquad S_{\psi \phi}=\sin(2|\beta_s|-2\varphi)
\ee

We note then:
\begin{itemize}
\item
While $\varepsilon_K$ can only be enhanced in this scenario, the fact that 
$b$ is a complex number implies the possibility of $|S(B_q)|$ being larger or
smaller than $S_{\rm SM}$, and therefore allows  for both enhancements and suppressions of $\Delta M_{s,d}$, independently of $\varepsilon_K$. In this manner  some tensions in the unitarity triangle fits can be avoided \cite{Barbieri:2011ci,Barbieri:2012uh}.
\item
In the $B_{s,d}$ meson systems  the suppressions of branching ratios are favoured 
by the $B_s\to\mu^+\mu^-$ data.
\item
Due to the measured value of $S_{\psi\phi}$ being SM-like, also the size of allowed modifications in $S_{\psi K_S}$ is  predicted to be small. As seen in (\ref{CPASYM}) the modifications of these two asymmetries are anti-correlated with each 
other and for fixed $\gamma$ this anti-correlation depends on the value of $\vub$ \cite{Buras:2012sd}. Similarly to MFV, this scenario favours then $\vub$ from exclusive decays, although it still allows for visible non-MFV effects.
\item
However, due to the breakdown of the correlation between $B_{s,d}$ and $K$ meson 
system, NP effects in $\kpn$ and $\klpn$ can be larger than in the MFV case, being only 
subject to constraints from $\varepsilon_K$, $\Delta M_K$, $K_L\to \mu^+\mu^-$ and also $\epe$. As we will see below the absence of correlation with 
$B_s\to \mu^+\mu^-$ is important here.
\end{itemize}


\boldmath
\subsection{$Z^\prime$ models with flavour symmetries}
\unboldmath

These models are less restrictive, and in the MFV case have 
four new real parameters relative to the SM,
\be
a, \qquad M_{Z^\prime},\qquad \Delta_L^{\nu\bar\nu}(Z^\prime),\qquad \Delta_A^{\mu\bar\mu}(Z^\prime),
\ee
where $\Delta(Z')$ denote the $Z'$ couplings to fermions,
and this number is reduced in the correlations between various observables. In 
the case of $U(2)^3$ models an additional complex parameter $b\not=a$ in 
$B_{s,d}$ systems is present and 
the correlations between the $K$ system and the $B_{s,d}$ systems are broken.

The relevant formulae for the shifts in various functions are obtained from 
the ones in the $Z$ models by simply 
replacing  $M_Z$  by $M_{Z^\prime}$ and the $Z$ couplings by $Z^\prime$ ones. The QCD correction $\tilde r$ in (\ref{DeltaS}) depends logarithmically on the $Z^\prime$ mass \cite{Buras:2012jb}. For definiteness we will set $\tilde r=0.941$, which corresponds to $M_{Z^\prime} = 5\tev$.

The crucial difference between $Z^\prime$ and $Z$ models is not only the big difference in their masses but more importantly that the $Z^\prime$ couplings to leptons 
are in principle arbitrary and do not have to satisfy the relation (\ref{LAZ}).
On the other hand, in accordance with the $SU(2)_L$ symmetry we have for all 
$Z^\prime$ models, independently of whether a flavour symmetry is imposed,
\be\label{SU2}
\Delta_{L}^{\nu\bar\nu}(Z^\prime) =\Delta_{L}^{\mu\bar\mu}(Z^\prime), \qquad 
\Delta_{V}^{\mu\bar\mu}(Z^\prime) = 2\Delta_L^{\nu\bar\nu}(Z^\prime) + \Delta_A^{\mu\bar\mu}(Z^\prime).
\ee
This will have interesting consequences as we will see below.
Moreover, these couplings and $M_{Z^\prime}$ are constrained by LEP II and 
present LHC data.
 
The correlations between various loop functions in the MFV case have now the
structure
\begin{align}
\Delta X&= \pm \frac{\sqrt{\Delta S^*}}{2\sqrt{\tilde r}}\frac{\Delta_L^{\nu\bar\nu}(Z^\prime)}{M_{Z^\prime}g_{\text{SM}}},\label{RRX}\\
\Delta Y&= \pm \frac{\sqrt{\Delta S^*}}{2\sqrt{\tilde r}}\frac{\Delta_A^{\mu\bar\mu}(Z^\prime)}{M_{Z^\prime}g_{\text{SM}}},\label{RRY}
\end{align}
and generally, in contrast to (\ref{RR1}), $\Delta X\not=\Delta Y$.
In the $U(2)^3$ scenario these formulae apply separately for the loop functions of the $K$ and $B_{s,d}$ systems, which generally differ from each other. Notice that $\Delta S$ is always real in the $U(3)^3$ case.

The following new features relative to the case of $Z$ models should be noted
\begin{itemize}
\item
As now $\Delta_L^{\nu\bar\nu}(Z^\prime)$ and $\Delta_A^{\mu\bar\mu}(Z^\prime)$ 
can differ from each other, the correlations between decays with muons and neutrinos in the final state are in principle absent. Therefore even in the MFV scenario the 
data on $B_s\to\mu^+\mu^-$ alone, being sensitive only to $\Delta_A^{\mu\bar\mu}(Z^\prime)$  have no impact on $\kpn$, $\klpn$ and $b\to s \nu\bar\nu$ transitions. However, when the data on $B\to K(K^*) \mu^+\mu^-$ are taken into 
account and the coupling $\Delta_{V}^{\mu\bar\mu}(Z^\prime)$ is restricted, 
the $SU(2)_L$ relation in (\ref{SU2}) implies some bounds on $\kpn$ and $\klpn$ 
in addition to those following from the allowed size of $\Delta S$. We will 
be more explicit about this issue in section~\ref{Anomalies} below.
\item
After $\Delta S$ and $b\to s \mu^+\mu^-$  constraints have been imposed, for fixed leptonic $Z^\prime$ couplings, NP effects in rare decays decrease with increasing $M_{Z^\prime}$ and 
as we will see in section~\ref{sec:6} for  $M_{Z^\prime}\ge 5\tev$ they will be 
rather small, in particular smaller than in particle-antiparticle mixing. 
This opposite hierarchy between NP effects in mixing and rare decays relative to the $Z$ case could allow one
in the future to distinguish $Z$ and $Z^\prime$ scenarios.
\item
In the $U(2)^3$ scenario also the correlations  between NP effects in 
$K\to\pi\nu\bar\nu$ decays and $B_{s,d}$ meson systems are broken allowing 
still significant enhancements of both branching ratios subject to the 
constraints from $\varepsilon_K$, $\Delta M_K$ and the LEP and LHC bounds 
on the $Z^\prime$ mass and its leptonic couplings.
\end{itemize}


\boldmath
\subsection{$Z$ and $Z^\prime$ with arbitrary FCNC quark couplings}
\unboldmath

Finally, we will investigate the cases of 
general FCNC quark couplings of $Z$ and $Z^\prime$ so that non-minimal sources of flavour violation will be present in all meson systems and generally they will not be correlated with each other. This will allow larger NP contributions to $\kpn$ and $\klpn$ than what was possible in the previous cases.

The simplest scenario of NP with non-minimal sources of flavour violation is 
the case of the $Z$ boson with FCNCs.
The only freedom in the kaon system in this NP scenario are the complex couplings $\Delta^{sd}_{L,R}(Z)$ as the $Z$ mass and its couplings to leptons are 
known. In $Z^\prime$  models, in addition to $\Delta^{sd}_{L,R}(Z^\prime)$,
two new real parameters enter: $M_{Z^\prime}$ and $\Delta_L^{\nu\bar\nu}(Z^\prime)$. 
In the latter case we will be guided by the bounds on the $Z^\prime$ mass and 
its leptonic couplings from LEP II and the LHC as well as LHCb data on 
$b\to s \mu^+\mu^-$ transitions.

These scenarios have already been considered in \cite{Buras:2012jb,Buras:2013qja,Buras:2014sba} but the treatment of CKM parameters was different there, and both the input 
from lattice QCD and the value of $\vcb$ have changed in the meantime.


\boldmath
\section{$\epe$}\label{sec:3a}
\unboldmath

\subsection{General Structure}

Let us begin our presentation of $\epe$ with the general formula for the effective Hamiltonian relevant for $K\to\pi\pi$ decays in any extension of the SM
\be\label{general}
\mathcal{H}_{\rm eff}(K\to\pi\pi)= \mathcal{H}_{\rm eff}(K\to\pi\pi)({\rm SM})+ 
\mathcal{H}_{\rm eff}(K\to\pi\pi)({\rm NP})
\ee
where  the SM part is given by 
\be\label{basic}
 \mathcal{H}_{\rm eff}(K\to\pi\pi)({\rm SM})=\sum_{i=1}^{10} C_i^{\rm SM}(\mu) Q_i
\ee
and the NP part by \be\label{ZprimeA}
\mathcal{H}_{\rm eff}(K\to\pi\pi)({\rm NP})=\sum_{i=1}^{10}(C_i(\mu)Q_i+C_i^\prime(\mu) Q_i^\prime).
\ee
Explicit expressions for the operators $Q_i$  can be found in  \cite{Buras:1993dy}. For our discussion 
it will be sufficient to have expressions only for the dominant QCD-penguin 
and electroweak penguin operators:
\begin{align}
&\text{\sc QCD Penguins:} &\notag\\
&Q_5 = (\bar s d)_{V-A} \sum_{q=u,d,s,c,b}(\bar qq)_{V+A}, &
Q_6 &= (\bar s_{\alpha} d_{\beta})_{V-A}\sum_{q=u,d,s,c,b}
       (\bar q_{\beta} q_{\alpha})_{V+A},\label{O3}\\
&\text{\sc Electroweak Penguins:} &\notag\\
&Q_7 = \frac{3}{2}\;(\bar s d)_{V-A}\sum_{q=u,d,s,c,b}e_q\;(\bar qq)_{V+A}, &
Q_8 &= \frac{3}{2}\;(\bar s_{\alpha} d_{\beta})_{V-A}\sum_{q=u,d,s,c,b}e_q
        (\bar q_{\beta} q_{\alpha})_{V+A}\label{O4}.
\end{align}
Here, $\alpha,\beta$ denote colours and $e_q$ denotes the electric quark charges reflecting the
electroweak origin of $Q_7,\ldots,Q_{10}$. Finally,
$(\bar qq')_{V\pm A}\equiv \bar q_\alpha\gamma_\mu(1\pm\gamma_5) q'_\alpha$. The so-called primed operators $Q_i^\prime$ 
are obtained from $Q_i$ by interchanging $V-A$ and $V+A$: these new operators 
contribute in the presence of right-handed flavour-violating couplings. Note that if NP scales are well above $m_t$, as is the case of $Z^\prime$ models,  the summation over 
flavours in (\ref{O3}) and (\ref{O4}) has to include also the top quark. But 
in the SM and $Z$ models the top quark is already integrated out.

The Wilson coefficients $C^{\rm SM}_i(\mu)$ 
are known at the NLO level in the renormalisation group improved perturbation theory including both QCD and QED corrections  \cite{Buras:1993dy,Ciuchini:1993vr}. Also some elements of NNLO corrections can be found in the literature \cite{Buras:1999st,Gorbahn:2004my}. 

If new operators beyond those present in the SM contribute to $\epe$ one 
should in principle perform the full RG analysis at the NLO level including these operators.
However, in view of various parameters involved  we will follow the procedure proposed in \cite{Buras:2014sba}  and consider NP contributions  at the LO. 
Moreover, as demonstrated there, at the end it is a good approximation to 
include in $\epe$ only  
 the modifications in the contributions of the dominant QCD penguin ($Q_6$)
and electroweak $(Q_8)$ operators and in the contribution of  the corresponding primed operators.

Now, relative to the case of $\kpn$, $\klpn$ and $\Delta F=2$  processes, 
flavour diagonal quark couplings are involved, and without knowing these couplings 
the correlation between rare $K$ decays and $\epe$ is lost. In the case of $Z$ 
the diagonal quark couplings are known and this implies a correlation between rare $K$ decays and $\epe$, as first stressed in  \cite{Buras:1998ed}.
But the case of $Z^\prime$ is different.
 For instance it 
could be that for some reason the flavour-diagonal quark couplings to $Z^\prime$ are very strongly 
suppressed relatively to the non-diagonal ones. In this case one 
would be able to enhance the branching ratios for  $\kpn$ and $\klpn$  without 
violating the $\epe$ constraint. We stress this point as the usual statements 
about correlation between rare $K$ decays and $\epe$  made in the literature apply to concrete models and one cannot exclude that through particular choices of flavour-diagonal $Z^\prime$ couplings to quarks this correlation can be broken. In what follows 
we will restrict our discussion to cases for which such correlations are present.

Finally, although the impact 
of $\epe$ also depends on the different scenarios for $Z$ couplings, as shown in 
\cite{Buras:2014sba}, the SM value of 
$\varepsilon_K$ must be consistent with the data if one wants to  satisfy simultaneously 
$\varepsilon_K$ and $\epe$. The details depend on 
the value of the hadronic matrix element of the QCD penguin operator $Q_6$, or 
equivalently on the value of the parameter $\bsi$. If $\varepsilon_K$ in 
the SM differs significantly from the data,  NP required to fit the data on 
$\varepsilon_K$ automatically violates the $\epe$ constraint for 
$\bsi$ within $20\%$ from its large $N$ value $\bsi=1.0$.
But, as we  shall see in detail in section~\ref{sec:lat}, a new insight in the 
 range of values of $\bsi$ has been gained through the studies in 
\cite{Buras:2015yba,Buras:2015xba}, so that now more space is left for NP contributions to $\epe$.
Also, as already mentioned, significant arbitrariness in the diagonal quark couplings 
to $Z^\prime$ allows for larger NP effects in this case.

In \cite{Buras:2015qea} we have updated the analysis of $\epe$ within the SM 
and the recent analyses of $\epe$ within $Z(Z^\prime)$ and 331 models 
have been presented in \cite{Buras:2014sba} and \cite{Buras:2014yna}, respectively. However, since then two improved analyses of $\epe$ in the SM have been 
presented \cite{Buras:2015yba,Buras:2015xba} and we will base our analysis on these two papers.


\subsection{SM Contribution}

The starting point of our presentation is the  analytic formula for $\epe$ within the SM \cite{Buras:2003zz,Buras:2014sba}, which has been recently 
updated in \cite{Buras:2015yba} and is given as follows 
\begin{equation}
\left(\frac{\varepsilon'}{\varepsilon}\right)_{\rm SM}= {\rm Im}\left[\lambda_t
F_{\varepsilon'}(x_t)\right],
\label{epeth0}
\end{equation}
 where
\begin{equation}
F_{\varepsilon'}(x_t) =P_0 + P_X \, X_0(x_t) + 
P_Y \, Y_0(x_t) + P_Z \, Z_0(x_t)+ P_E \, E_0(x_t)~.
\label{FE0}
\end{equation}
The first term in (\ref{FE0}) is dominated by QCD-penguin contributions, the next three 
terms by electroweak penguin contributions and the last term is
totally negligible. 
The $x_t$ dependent functions have been 
collected in  the Appendix~A of \cite{Buras:2015qea}.

The coefficients $P_i$ are given in terms of the non-perturbative parameters\footnote{Note that $R_i$ do not contain the factor $1.13$ present in \cite{Buras:2014sba}.}
\be\label{RS}
R_6\equiv \bsi\left[ \frac{114.54\mev}{m_s(m_c)+m_d(m_c)} \right]^2,
\qquad
R_8\equiv \bei\left[ \frac{114.54\mev}{m_s(m_c)+m_d(m_c)} \right]^2,
\ee
 as follows:
\begin{equation}
P_i = r_i^{(0)} + 
r_i^{(6)} R_6 + r_i^{(8)} R_8 \,.
\label{eq:pbePi}
\end{equation}
The coefficients $r_i^{(0)}$, $r_i^{(6)}$ and $r_i^{(8)}$ comprise
information on the Wilson-coefficient functions of the $\Delta S=1$ weak
effective Hamiltonian at the NLO. Their numerical values  for 
three  values of $\alpha_s(M_Z)$ are collected 
in the Appendix~B of \cite{Buras:2015yba}.
We will next describe how the (\ref{epeth0}) is modified in the presence of NP 
contributions. The structure of modifications depends on NP model considered.

In our numerical analysis 
we will use for the quark masses the values of \cite{Aoki:2013ldr}, given in table~\ref{tab:input}.
Then at the nominal value $\mu=m_c=1.3\gev$ we have 
\be
m_s(m_c)=(109.1\pm2.8) \mev, \qquad
m_d(m_c)=(5.44\pm 0.19)\mev.
\ee


\boldmath
\subsection{CMFV and $U(2)^3$}
\unboldmath

These are the simplest cases as only the shifts in the function $X,Y,Z$, discussed 
in previous section, have to be made if NP is not far from the electroweak scale.
The most predictive in this case is $Z$ scenario as in this case
the following  shifts in 
the functions $X$, $Y$ and $Z$ entering the 
analytic formula (\ref{epeth0}) have to be made 
\be\label{eprimeshiftsCMFV}
\Delta X=\Delta Y =\Delta Z= a c_W\frac{8\pi^2}{g^3}\, ,
\ee 
which equal just the shifts in (\ref{XSHIFT}).
The reason why the shift is universal in these three functions originates in 
the fact that a $Z$ exchange with flavour violating couplings in one vertex 
and known flavour diagonal couplings modifies just the $Z$-penguin contribution 
which universally enters $X$, $Y$ and $Z$.

The shift $\Delta Z$ has the largest impact on $\epe$, as the coefficient 
$P_Z$ is large and negative. For a positive $a$ the enhancement of $\kpn$ and 
$\klpn$ implies suppression of $\epe$, while a negative $a$ suppresses these 
branching ratios and enhances $\epe$. In fact, in MFV this scenario appears to be favoured by the $B_s\to\mu^+\mu^-$ data. Moreover, it would also 
be favoured by the data on $\epe$, if the SM prediction for $\epe$ will turn out to be below its measured value, as presently indicated by the analyses in 
\cite{Bai:2015nea,Buras:2015yba,Buras:2015xba}

The correlation with $\B(\klpn)$ is made manifest using expression \eqref{ImX} together with \eqref{eprimeshiftsCMFV}, from which one has
\begin{equation}\label{epsilonprimeKlpn}
\left(\frac{\epsilon'}{\epsilon}\right)_{\rm MFV} = \left(\frac{\epsilon'}{\epsilon}\right)_{\rm SM} + h\,
(P_X + P_Y + P_Z)\left[\lambda^5\left(\frac{\B(\klpn)}{\kappa_L}\right)^{1/2} - X_{\rm SM}{\rm Im}\,\lambda_t\right],
\end{equation}
while the correlation with $\B(\kpn)$ follows from the fact that the phase of $X_{\rm eff}$ in \eqref{ReX} is aligned with the SM.

If the flavour symmetry is reduced down to $U(2)^3$ the formula in 
(\ref{eprimeshiftsCMFV}) is still valid but the correlation with $B_{s,d}$ 
meson systems is broken and the constraints on the NP contributions to $\epe$ are weaker. In particular, independently of $B_s\to\mu^+\mu^-$, the ratio $\epe$ can be 
enhanced or suppressed but its MFV correlation with $\kpn$ and $\klpn$ remains 
valid.

The case of $Z^\prime$ is complicated by the fact that the diagonal quark 
couplings are rather arbitrary,  and are not constrained by other semileptonic rare decays. The analysis of $\epe$ can therefore not be very specific even if constraints from LEP and LHC are taken into account. 
It should also be emphasized that, depending on the structure of diagonal couplings, different operators dominate $\epe$ (even if generally they are $Q_6$, $Q_8$ or the corresponding primed operators).   The 
good news in $Z^\prime$ scenarios is that unless a concrete framework is 
considered, there is no strict correlation between $\epe$ and $K\to\pi\nu\bar\nu$ 
allowing for larger NP effects in these decays than what is possible in the 
case of $Z$ scenarios.


\boldmath
\subsection{$Z$ with  general flavour-violating couplings}
\unboldmath

It should be emphasized, that this scenario can be realized in many models 
and in the case of the absence of a discovery of new particles at the LHC 
the flavour violating couplings of $Z$ could constitute an important window to short distance scales beyond the LHC.

For completeness we recall here the formulae for $\epe$ derived in \cite{Buras:2014sba}. The details including derivations can be found there.
Relative to the $Z^\prime$ case, discussed subsequently, the RG  running in this case is simplified by the fact that the initial conditions for the Wilson coefficients have to be evaluated at the electroweak scale as in the SM. We consider three scenarios for the quark couplings: only left-handed (LH), only right-handed (RH), and left-right symmetric (LRS) \cite{Buras:2012jb}. In the 
ALRS scenario of \cite{Buras:2012jb} the NP contributions to $\klpn$ and $\kpn$ vanish and this case is 
uninteresting from the point of view of the present paper.

\subsubsection{LH scenario}
Here the simplest approach is to make the following shifts in 
the functions $X$, $Y$ and $Z$ entering the 
analytic formula \eqref{epeth0} \cite{Buras:2014sba}:
\begin{equation}\label{eprimeshifts}
\Delta X=\Delta Y =\Delta Z= c_W\frac{8\pi^2}{g^3}\frac{\Delta_L^{sd}(Z)}{\lambda_t}\, .
\end{equation}
This formula gives the generalization of the shifts in 
\eqref{eprimeshiftsCMFV} to arbitrary LH 
flavour-violating $Z$ couplings to quarks.
We have then
\begin{equation}\label{epsLHS}
\left(\frac{\varepsilon'}{\varepsilon}\right)_\text{LHS}=
\left(\frac{\varepsilon'}{\varepsilon}\right)_\text{SM}+
\left(\frac{\varepsilon'}{\varepsilon}\right)_{Z,L}
\end{equation}
where the second term stands for the modification related to \eqref{eprimeshifts}.

Since the shifts in the loop functions \eqref{eprimeshifts} are universal, the correlation between $\epsilon'$ and $\B(\klpn)$ is again given by \eqref{epsilonprimeKlpn}. On the other hand, since the phase of the $\Delta_L^{sd}$ coupling is now arbitrary, the correlation with $\B(\kpn)$ is lost in this case.

\subsubsection{RH scenario}
This case is analyzed in detail in section 7.5 in  \cite{Buras:2014sba}, 
where it is demonstrated that by far the dominant new contribution to $\epe$ 
comes from the  $Q^\prime_8$ operator. The relevant hadronic matrix element 
$\langle Q^\prime_8\rangle_2=-\langle Q_8\rangle_2$ and consequently   it is known from lattice QCD \cite{Blum:2012uk,Blum:2015ywa}. We refer to \cite{Buras:2014sba} for details.

In this case we have then
\be\label{epsRHS}
\left(\frac{\varepsilon'}{\varepsilon}\right)_\text{RHS}=
\left(\frac{\varepsilon'}{\varepsilon}\right)_\text{SM}+
\left(\frac{\varepsilon'}{\varepsilon}\right)_{Z,R}\, 
\ee
with the second term given  within an excellent approximation by \cite{Buras:2014sba}
\be\label{eprimeZfinal}
\left(\frac{\varepsilon'}{\varepsilon}\right)_{Z,R}=
-6.2 \times 10^3 \,  \left[\frac{114\mev}{m_s(m_c) + m_d(m_c)}\right]^2 \,
\left[\frac{B_8^{(3/2)}}{0.76}\right]\,{\rm Im}\,\Delta_R^{sd}(Z)\,.
\ee
Note that due to the new lattice results in \cite{Blum:2015ywa} the central value of $\bei$ has been modified relative to \cite{Buras:2014sba} where the older 
value $0.65$ extracted from \cite{Blum:2012uk} has been used.

This result implies that ${\rm Im}\,\Delta_R^{sd}(Z)$ must be at most be $\ord(10^{-7})$ in order for 
$\epe$ to agree with experiment. Then, similarly to the LH case just 
discussed, NP contribution to $\varepsilon_K$ are very small and only for 
CKM parameters for which $\varepsilon_K$ in the SM agrees well with the 
data this scenario remains viable.

As far as $\kpn$ and $\klpn$ are concerned we can use the formulae in 
\cite{Buras:2012jb}. Equivalently, in the case of the RH scenario, one can just 
make a shift in the function $X(K)$:
\be\label{XLKZ}
\Delta X(K)=\left[\frac{\Delta_L^{\nu\bar\nu}(Z)}{g^2_{\rm SM}M_{Z}^2}\right]
             \left[\frac{\Delta_R^{sd}(Z)}{\lambda_t}\right], \qquad 
\Delta_L^{\nu\bar\nu}(Z)=\frac{g}{2c_W}.
\ee
Expressing ${\rm Im}\,\Delta_R^{sd}$ in terms of $\B(\klpn)$ through \eqref{ImX}, one then has
\begin{equation}
\left(\frac{\epsilon'}{\epsilon}\right)_{Z,R} = -32.6\cdot R_8\, \left[\lambda^5\left(\frac{\B(\klpn)}{\kappa_L}\right)^{1/2} - {\rm Im}\,\lambda_t\cdot X_{\rm SM}\right].
\end{equation}

\subsubsection{General case}
When both $\Delta_L^{sd}(Z)$ and $\Delta_R^{sd}(Z)$ are present
the general formula for $\epe$ is
\be\label{epsgeneral}
\left(\frac{\varepsilon'}{\varepsilon}\right)=
\left(\frac{\varepsilon'}{\varepsilon}\right)_\text{SM}+
\left(\frac{\varepsilon'}{\varepsilon}\right)^L_{Z}+
\left(\frac{\varepsilon'}{\varepsilon}\right)^R_{Z}
\ee
with the last two terms representing LH and RH contributions discussed 
above. This formula allows to calculate $\epe$ for arbitrary $Z$ couplings, in particular for the LRS scenario where $\Delta_L^{sd}(Z)=\Delta_R^{sd}(Z)$, and for the case presented in section~\ref{ZmodelLR}.

The numerical analysis of all these scenarios is presented in section~\ref{sec:6}.


\boldmath
\subsection{$Z^\prime$ with  flavour-violating couplings}
\unboldmath

We have already emphasized that in general, in the absence of the knowledge 
of flavour diagonal $Z^\prime$ couplings to quarks, there is no correlation 
between $\epe$ and $K\to\pi\nu\bar\nu$ decays. We will therefore not present 
a numerical analysis of $\epe$ in $Z^\prime$ scenarios, except for one case in section~\ref{sec:lat}.

The analysis  in 331 models, where the operator $Q_8$ turns 
out to be most important, can be found in \cite{Buras:2014yna}. On the other hand, in \cite{Buras:2014sba}, where the possible impact of $Z^\prime$ on the $\Delta I=1/2$ rule 
has been considered, the diagonal couplings could be fixed 
 by requiring the maximal 
contribution of $Z^\prime$ to the $A_0(K\to\pi\pi)$ amplitude. In this case 
the operator $Q_6$ turned out to be most important. As we will see below, a variant of this model turns out to be interesting in 
view of the recent lattice result on $\epe$ in \cite{Bai:2015nea} and recent analyses in \cite{Buras:2015yba,Buras:2015xba}.


\boldmath
\subsection{Can $\epe$ and $K\to\pi\nu\bar\nu$ be simultaneously enhanced?}\label{sec:lat}
\unboldmath

In most extensions of the SM found in the literature the enhancement of the branching ratio for $\klpn$
through NP usually implies the suppression of $\epe$, and vice versa an enhancement of $\epe$ implies a suppression of $\klpn$. We have already 
mentioned this feature in the context of our analysis of $Z$ models with MFV 
after \eqref{eprimeshiftsCMFV}.
This is related
to the fact that there is a strong correlation between the {\it negative} electroweak penguin contribution to $\epe$ and the branching ratio for $\klpn$. Here we would like to present two simplified models in which in fact 
$\epe$ and $\B(\klpn)$ can be simultaneously 
enhanced with respect to their SM values.

This case is of interest in view of the recent result from the RCB-UKQCD 
lattice collaboration which indicates that $\epe$ in the SM could be significantly below the data. Indeed, they find 
in the SM \cite{Bai:2015nea}
\be\label{RBC}
(\epe)_\text{SM}=(1.4 \pm 7.0)\cdot 10^{-4}~,
\ee
which is by $2.1\sigma$ below the experimental world average from the
NA48 \cite{Batley:2002gn}   and 
KTeV \cite{AlaviHarati:2002ye,Worcester:2009qt} collaborations, 
\be\label{EXP}
(\epe)_\text{exp}=(16.6\pm 2.3)\cdot 10^{-4}\,.
\ee

A recent detailed anatomy of $\epe$ in the SM in \cite{Buras:2015yba} also confirms that, with the value of $\bsi$ from \cite{Bai:2015nea}, $\epe$ in the SM 
 is indeed significantly smaller than the experimental value. Assuming that the real parts of the 
$K\to\pi\pi$ amplitudes are fully governed by the SM dynamics and including 
isospin breaking effects the authors of \cite{Buras:2015yba} find 
\be\label{BGJJ}
(\epe)_\text{SM}=(1.9 \pm 4.5)\cdot 10^{-4}~,
\ee
which is by $2.9\sigma$ below \eqref{EXP}. Clearly, 
the size of this suppression of $\epe$  depends sensitively on the 
value of $\bsi$,  the dominant source of uncertainty in the prediction of $\epe$ in the 
SM. But even discarding lattice results, and using the recently derived upper bounds 
on $\bsi$ and $\bei$ from the large $N$ approach \cite{Buras:2015xba}, $\epe$ is 
found typically by a factor of two below the data.
Motivated by these finding we looked for models in which $\epe$ and 
$\klpn$ could be simultaneously enhanced.

\boldmath
\subsubsection{Simplified $Z$ model}\label{ZmodelLR}
\unboldmath
We consider a model in which $Z$ has both LH and RH couplings, but not equal to 
each other, and not differing only by a sign. As seen in (\ref{eprimeZfinal}), 
in order to obtain a positive contribution to $\epe$ we need ${\rm Im}\,\Delta_R^{sd}(Z)<0$. But this alone would suppress the rare decay branching ratios. 
The solution to this problem is the contribution of the $Q_8$ operator to $\epe$ given in (\ref{eprimeshifts}). While this is not evident from this formula, as shown in \cite{Buras:2014sba}, for equal LH and 
RH $Z$ couplings this contribution is by a factor of $3.3$ smaller than the 
one in  (\ref{eprimeZfinal}). On the other hand, the branching ratio for
$\klpn$ is sensitive to the sum of LH and RH couplings. Therefore 
choosing ${\rm Im} \Delta_L^{sd}(Z)> 0$ with 
\be
|{\rm Im} \Delta_R^{sd}(Z)|<{\rm Im} \Delta_L^{sd}(Z)< 3.3 |{\rm Im} \Delta_R^{sd}(Z)|
\ee
one can enhance simultaneously $\epe$ and the branching ratio for $\klpn$. In doing this, ${\rm Re}\,\Delta_{L,R}^{sd}(Z)$ have to be kept sufficiently
 small in order not to spoil the agreement of ${\rm Re} A_0$ in the SM with
the data. Moreover, the $\Delta M_K$ and $\varepsilon_K$ constraints have to 
be satisfied.

\begin{figure}
\centering%
\includegraphics[width=0.32\textwidth]{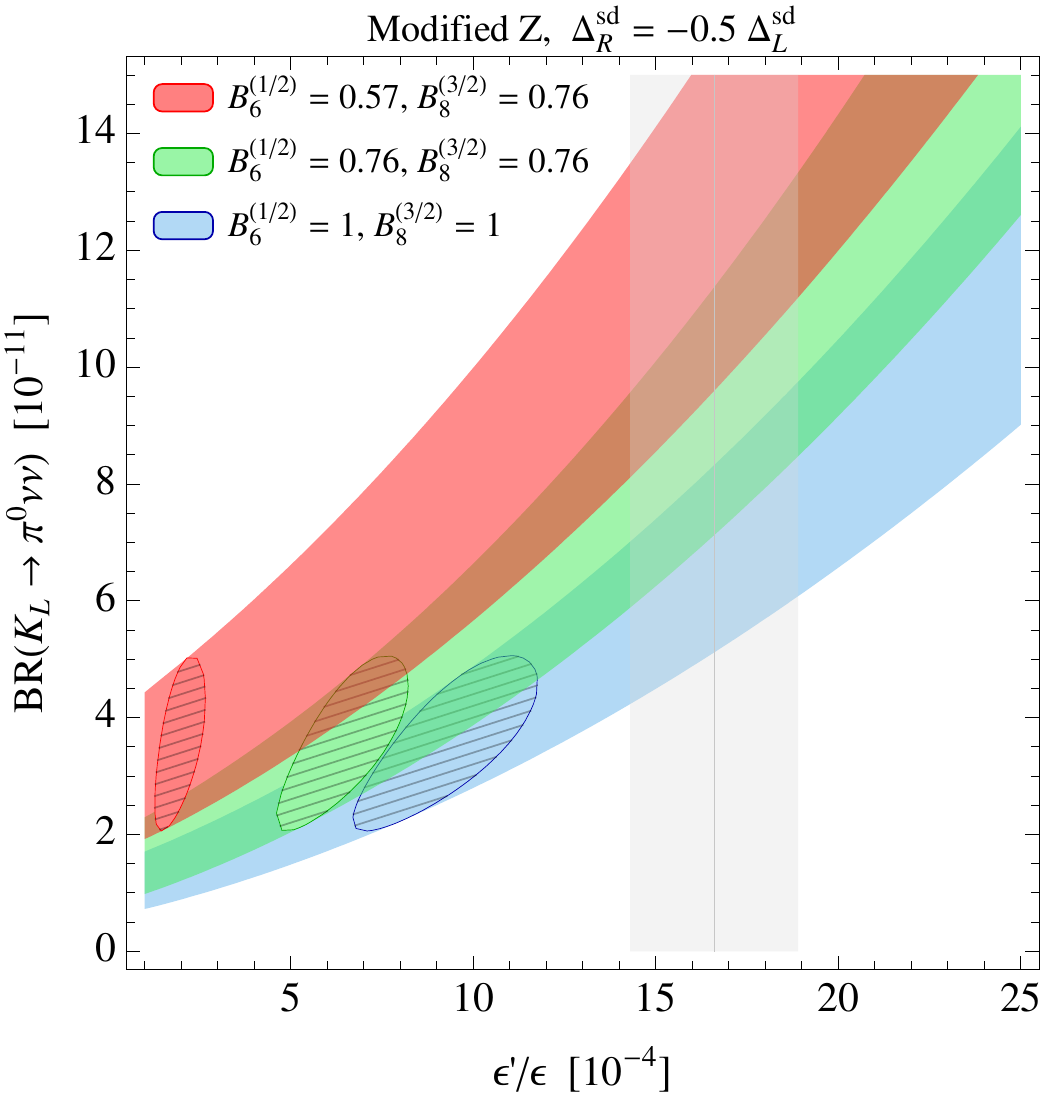}\hfill%
\includegraphics[width=0.32\textwidth]{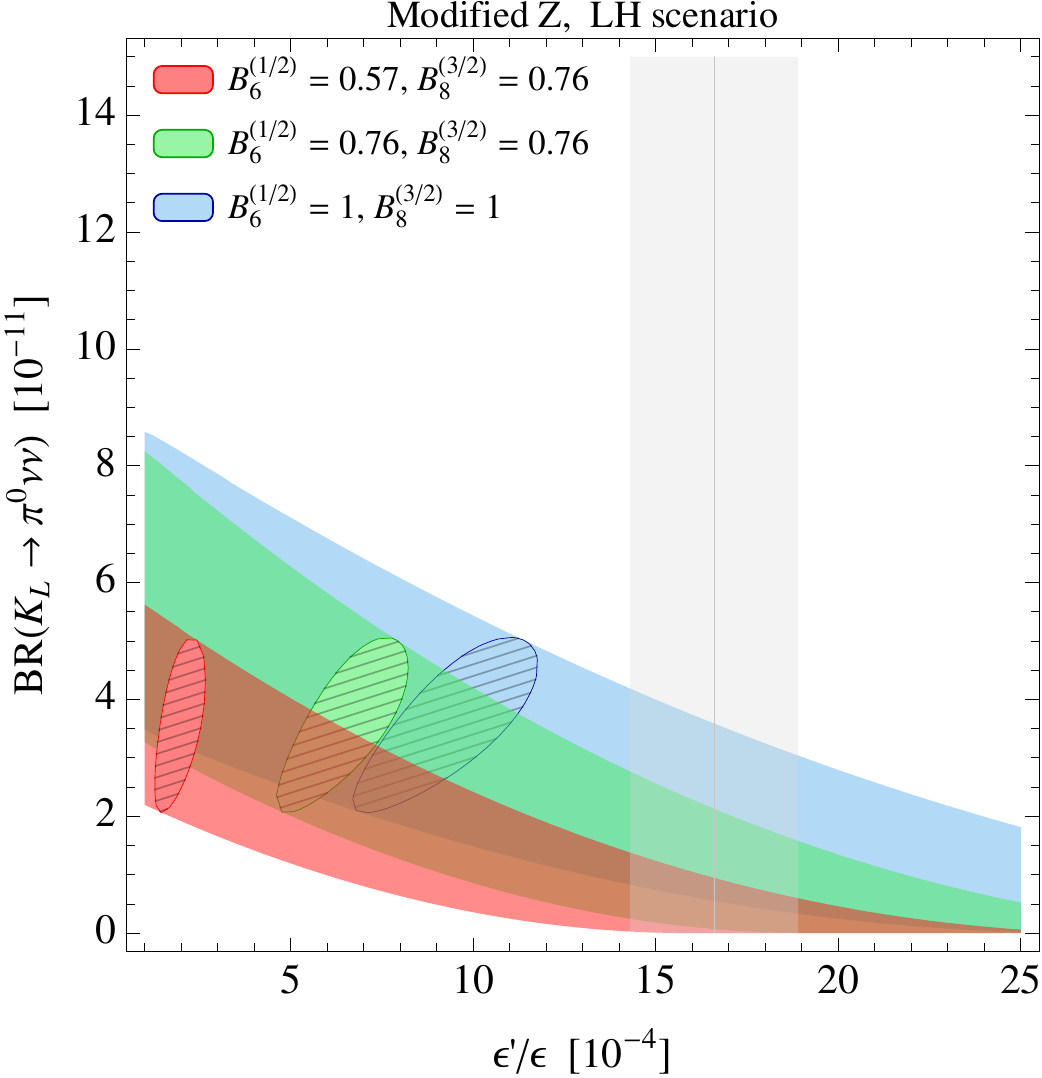}\hfill%
\includegraphics[width=0.32\textwidth]{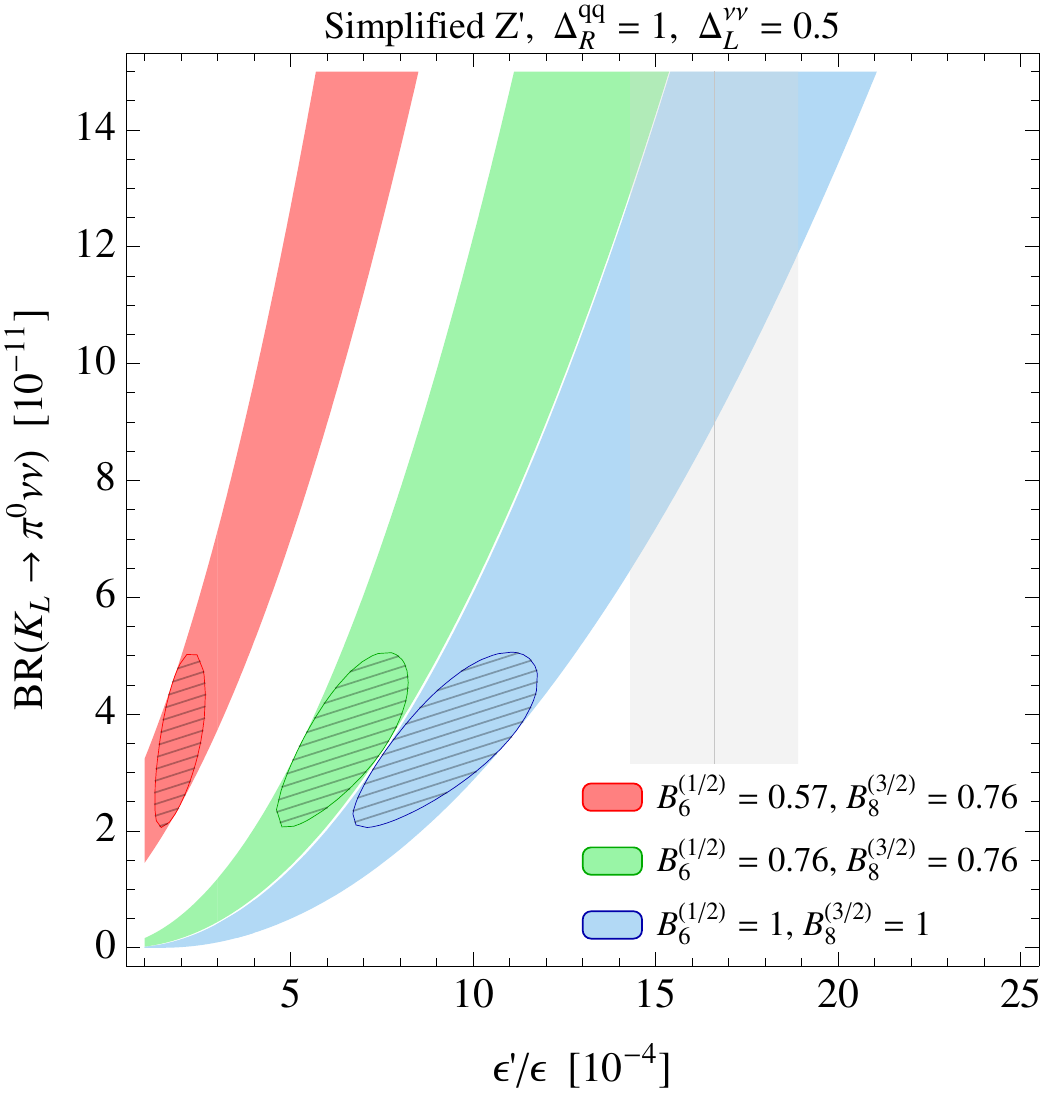}
\caption{\it 95\% C.L. allowed regions for $\epe$ and $\klpn$. Left: model with flavour-changing Z boson couplings $\Delta_R^{sd} = -0.5 \Delta_L^{sd}$. Center: modified Z, LH scenario $\Delta_R^{sd} = 0$. Right: 5 TeV Z' with $\Delta_R^{qq} = 1$ and $\Delta_L^{\nu\nu} = 0.5$. The plots are for $B_6 = 1$ (blue), $B_6 = 0.76$ (green), and $B_6 = 0.57$ (red). The hatched regions are the SM predictions at $2\sigma$. The gray band shows the experimental result for $\epe$.\label{ZepsilonKL}}
\end{figure}

In the left panel of figure~\ref{ZepsilonKL} we show the correlation between $\epe$ and $\klpn$ in the case of $\Delta_L^{sd}(Z) = -2\Delta_R^{sd}(Z)$, and compare it with the opposite correlation that is present in the LH scenario (central panel). The different colours correspond to different choices of the parameters $\bsi$ and $\bei$:
\begin{align}
&& \bsi&=1.0, & \bei&=1.0 & &{(\text{blue})}, && \label{AJB1}\\
&& \bsi&=0.76, & \bei&=0.76 & &{(\text{green})}, && \label{AJB2}\\
&& \bsi&=0.57, & \bei&=0.76 & &{(\text{red})}\,. && \label{AJB3}
\end{align}
The first choice is motivated by the upper bound from large $N$ approach 
\cite{Buras:2015xba}, $\bsi\le\bei <1$.
The second choice uses the central value for $\bei$ from the RBC-UKQCD collaboration
 \cite{Blum:2015ywa} extracted in \cite{Buras:2015qea}, and assumes that 
$\bsi=\bei$ saturating the previous bound.
Finally, the third choice uses the
central values for both $\bsi$ and $\bei$ from the RBC-UKQCD collaboration, with 
$\bsi$ extracted in \cite{Buras:2015yba} from the lattice results in \cite{Bai:2015nea}.

As expected, in our simple model the requirement of satisfying the data on 
$\epe$ automatically implies enhanced values of $\mathcal{B}(\klpn)$, while 
in the LH model, similar to the Littlest Higgs model with T-parity \cite{Blanke:2015wba}, suppressed $\mathcal{B}(\klpn)$ is predicted.

We do not present the correlation between $\epe$ and $\kpn$ as this also
involves real parts of the new couplings and is more model dependent.

\boldmath
\subsubsection{Simplified $Z^\prime$ model}
\unboldmath
Another example of a model in which $\B(\klpn)$ and $\epe$ can be simultaneously enhanced has been already considered in 
\cite{Buras:2014sba}. In this model, not the electroweak penguin operator $Q_8$, 
but the QCD penguin operator $Q_6$ is affected by NP. A tree-level exchange 
of $Z^\prime$ with left-handed flavour violating quark couplings and flavour universal structure of diagonal RH quark couplings 
generates the $Q_5$ operator, and through renormalisation group evolution also the
$Q_6$ operator which at the end dominates the NP contribution to $\epe$. 

Assuming then that $Z^\prime$ has only LH flavour violating couplings one
has \cite{Buras:2014sba}
\be\label{IMNP}
{\IM} A_0^{\rm NP}= {\IM} C_6(\mu)\langle Q_6(\mu)\rangle_0,
\ee
where
\be\label{LOC5C6}
 C_6(m_c)= 1.13\frac{\Delta_L^{s d}(Z^\prime)\Delta_R^{qq}(Z^\prime)}{4 M^2_{Z^\prime}}.
\ee
and
\be\label{eq:Q60}
\langle Q_6(\mu) \rangle_0=-\,4 
\left[ \frac{m_{\rm K}^2}{m_s(\mu) + m_d(\mu)}\right]^2 (F_K-F_\pi)
\,B_6^{(1/2)}\,.
\ee
Clearly the size of the NP effects depend on the various couplings of the $Z'$ to quarks and leptons.
The right panel of figure~\ref{ZepsilonKL} shows the results for the values
\begin{align}\label{setting}
\Delta_R^{qq}(Z^\prime) &= 1, & \Delta_L^{\nu\nu}(Z^\prime) &= 0.5,
\end{align}
which satisfy the LHC bounds on flavour-conserving four-fermion interactions, and again for the three choices of the parameters $B_6^{(1/2)}$ and $B_8^{(3/2)}$ of \eqref{AJB1}--\eqref{AJB3}.


\boldmath
\section{Relations to other $\Delta F = 1$ processes}\label{sec:4}
\unboldmath 

\boldmath
\subsection{$b\to s \mu^+\mu$}\label{Anomalies}
\unboldmath

It is of interest to see how the decays $\kpn$ and $\klpn$ are correlated 
with $b\to s\mu^+\mu^-$ transitions and in particular what are the implications 
of the $B\to K(K^*)\mu^+\mu^-$ anomalies for  $\kpn$ and $\klpn$ in the context
of the simplest models.

Let us first note that $Z$ models of any kind cannot explain these anomalies 
for various reasons. In concrete models these anomalies are most easily explained 
through the shifts in the Wilson coefficients $C_9$ and $C_{10}$ of the 
operators
\begin{align}
Q_9 &= (\bar s\gamma_\mu P_L b)(\bar \ell\gamma^\mu\ell), &
Q_{10} &= (\bar s\gamma_\mu P_L b)(\bar \ell\gamma^\mu\gamma_5\ell)\,,
\end{align}
 with 
 \cite{Descotes-Genon:2014uoa,Altmannshofer:2014rta,Hiller:2014yaa,Altmannshofer:2015sma}
\be\label{910}
C_9^\text{NP}\approx  -C^\text{NP}_{10} \approx -(0.5\pm 0.2) \,.
\ee
The solution with NP 
present only in $C_9$, with $C_9^\text{NP}\approx -1$,  is even favoured, but much harder to explain in the context of  existing models. We refer to \cite{Altmannshofer:2015sma} for tables 
with various solutions.

\begin{figure}[t]
\centering%
\includegraphics[width=0.47\textwidth]{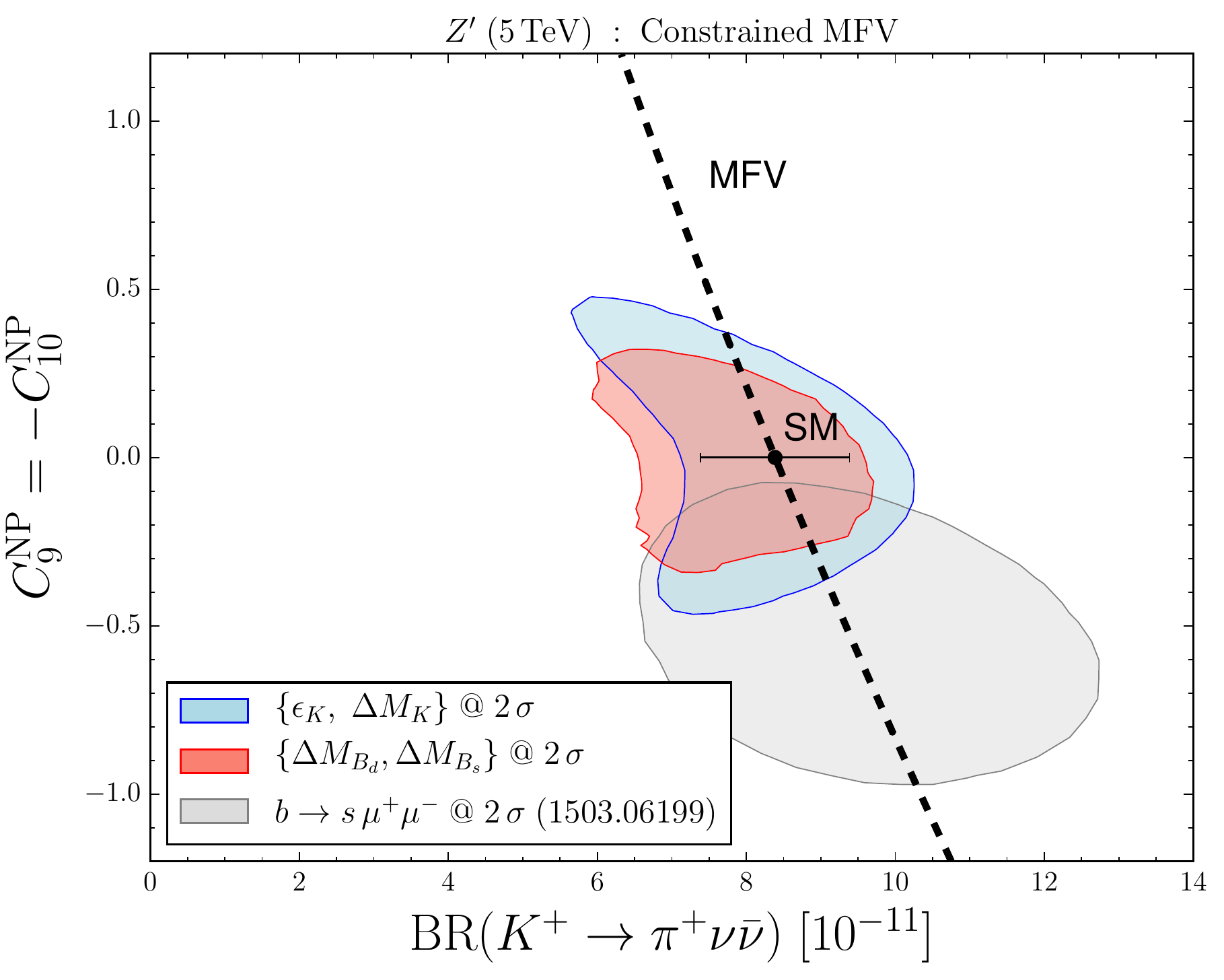}\hfill%
\includegraphics[width=0.47\textwidth]{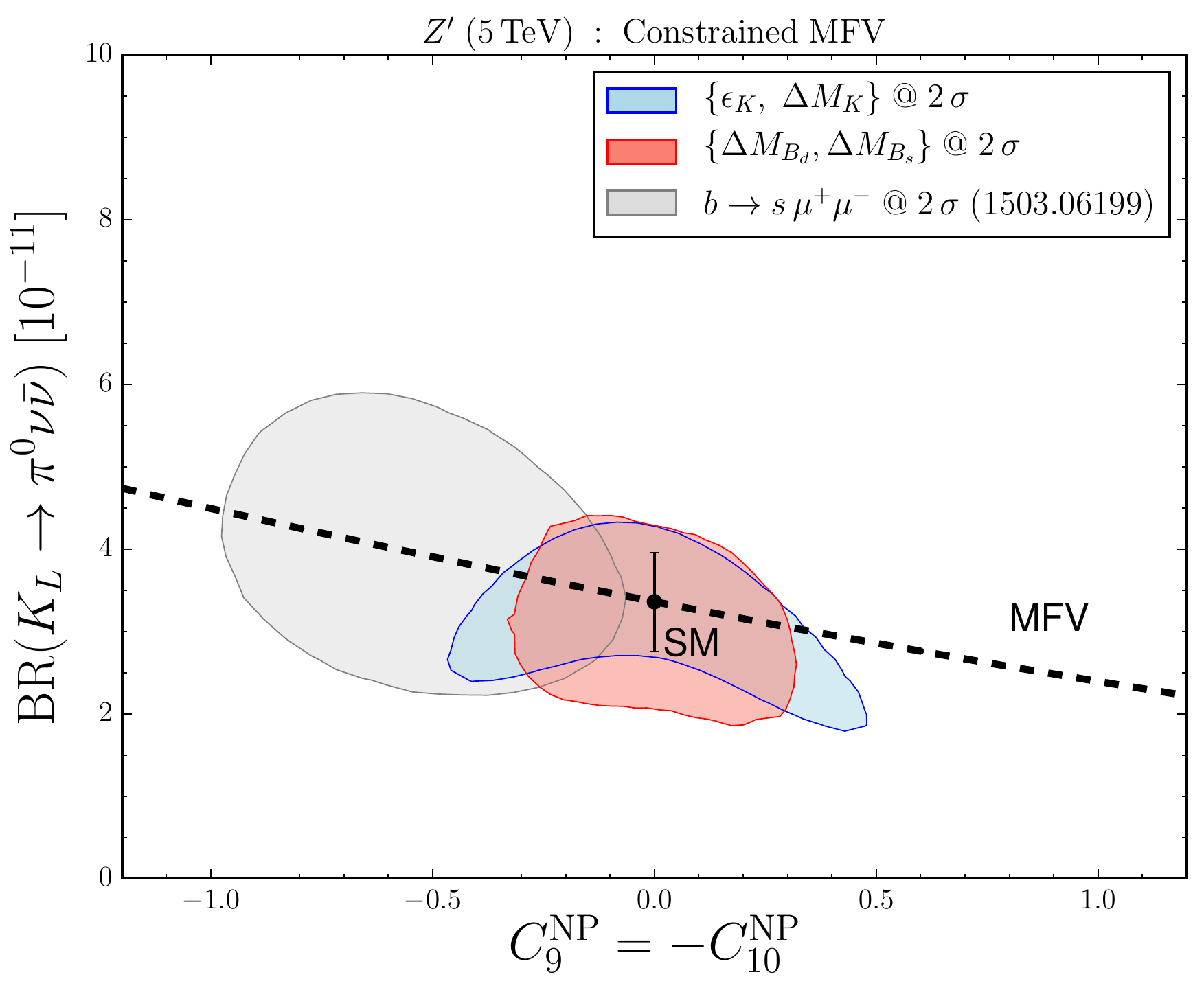}
\caption{\it Allowed ranges for $C_9^{\rm NP} = - C_{10}^{\rm NP}$ versus  $\mathcal{B}(\klpn)$ (left panel) and $\mathcal{B}(\kpn)$ (right panel) in a simplified $5\tev~Z'$ model obeying CMFV\@. The $2\,\sigma$ confidence regions shown correspond to constraints from kaon mixing (blue), $B$ mixing (red) and $b\to s\mu^+\mu^-$ transitions (grey) (from \cite{Altmannshofer:2015sma}).\label{fig:Knn-C9-CMFV}}
\end{figure}

This relation is very badly violated in $Z$ models for which one has
\be
\frac{C^\text{NP}_{10}}{C_9^\text{NP}}= \frac{\Delta_A^{\mu\bar\mu}(Z)}{\Delta_V^{\mu\bar\mu}(Z)}=-13.3
\ee
in drastic disagreement with (\ref{910}). 
The explanation of $B\to K^*\mu^+\mu^-$ anomalies  would then imply very strong suppression of $\mathcal{B}(B_s\to\mu^+\mu⁻)$ relative to the SM which disagrees with the data. On the other hand the agreement with the data on $\mathcal{B}(B_s\to\mu^+\mu⁻)$ would allow only very small value of $C_9^\text{NP}$.

In $Z^\prime$ models we have  generally
\be\label{C9}
\sin^2\theta_W C_9^\text{NP}=-\frac{\Delta_L^{sb}(Z^\prime)} {V_{ts}^* V_{tb}} \frac{\Delta_V^{\mu\bar\mu}(Z^\prime)}{M_{Z^\prime}^2g_{\text{SM}}^2}
\ee
\be\label{C10} 
\sin^2\theta_W C_{10}^\text{NP}=-\frac{\Delta_L^{sb}(Z^\prime)} {V_{ts}^* V_{tb}} \frac{\Delta_A^{\mu\bar\mu}(Z^\prime)}{M_{Z^\prime}^2g_{\text{SM}}^2}
\ee
Therefore for 
\be 
\Delta_V^{\mu\bar\mu}(Z^\prime)=-\Delta_A^{\mu\bar\mu}(Z^\prime)
\ee
the relation between $C_9^\text{NP}$ and $C^\text{NP}_{10}$ in (\ref{910}) can 
be satisfied. This is the case of $Z^\prime$ with purely $V-A$ couplings both in the quark and lepton sector.

But the $SU(2)_L$ relation in (\ref{SU2}) then implies that 
\be\label{LVZprime}
\Delta_L^{\nu\bar\nu}(Z^\prime)=\Delta_V^{\mu\bar\mu}(Z^\prime)\,.
\ee
In turn in the case of MFV, when the first ratio on the r.h.s in (\ref{C9}) and   (\ref{C10}) reduces to flavour independent $a$, we have 
\be
\Delta X_L(K)=\Delta X_L(B_d)=\Delta X_L(B_s)\equiv \Delta X= -\sin^2\theta_W C_9^\text{NP}.
\ee
\be
\Delta Y_L(K)=\Delta Y_L(B_d)=\Delta Y_L(B_s)\equiv \Delta Y= \sin^2\theta_W C_9^\text{NP}.
\ee

Therefore, for $Z^\prime$ models with MFV quark couplings, the $B\to K(K^*)\mu^+\mu^-$ anomalies imply:
\begin{itemize}
\item
Enhancement of the branching ratios  $\mathcal{B}(\kpn)$ and  $\mathcal{B}(\klpn)$ relative to their SM values;
\item
Suppression of the branching ratios  $\mathcal{B}(B_s\to \mu^+\mu^-)$ and 
 $\mathcal{B}(B_d\to \mu^+\mu^-)$ relative to their SM values;
\item
Enhancement of the branching ratios  $\mathcal{B}(B\to K^*\nu\bar\nu)$ and  
 $\mathcal{B}(B\to K \nu\bar\nu)$  relative to their SM values as already 
pointed out in \cite{Buras:2014fpa}.
\end{itemize}

The first of these results does not apply beyond MFV, even in $U(2)^3$ models, but the second and 
third remain true in  $U(2)^3$ models. Moreover,  for arbitrary $Z^\prime$ quark couplings the correlations  between  $B\to K(K^*)\mu^+\mu^-$,   $\mathcal{B}(B_s\to \mu^+\mu^-)$ and $\mathcal{B}(B\to K^*\nu\bar\nu)$ exist due to the $SU(2)_L$ relation in (\ref{SU2})
as already known from other analyses, in particular \cite{Buras:2014fpa}.
In the latter case we can compare the region still allowed for $5\,\tev$ $Z'$ shown in the right panel of figure~\ref{fig:Knn-CMFV} with the fit results on $C_9$ from \cite{Altmannshofer:2015sma}.

In figure~\ref{fig:Knn-C9-CMFV} we show the regions still allowed in the $C_9^{\rm NP} = - C_{10}^{\rm NP}$ versus  $\mathcal{B}(\klpn)$ and $\mathcal{B}(\kpn)$ planes, in a simplified $5\tev~Z'$ model obeying CMFV.

 We observe that for $C_9^{\rm NP}\le -0.3$ one leaves the $2\sigma$ range allowed by $\Delta M_{s,d}$, and for  $C_9^{\rm NP}\le -0.5$ the one allowed 
by $\varepsilon_K$ and $\Delta M_K$. Thus a massive $Z^\prime$ with MFV couplings
 can lower the tension of the theory with data but cannot fully explain the
observed anomaly.


\boldmath
\subsection{$B\to K(K^*)\nu\bar\nu$}\label{sec:BKnn}
\unboldmath

There are many reasons for performing an analysis of $B\to K^{(*)}\nu\bar\nu$ decays in our paper:
\begin{itemize}
\item
It is well known that they are strongly correlated with $K\to \pi \nu\bar\nu$ 
decays in models with MFV \cite{Buras:2001af}, but also in more complicated 
models \cite{Buras:2013ooa}.
\item
As recently shown in \cite{Buras:2014fpa} these decays, when measured, could allow to distinguish 
between various explanations of the present anomalies in $b\to s \mu^+\mu^-$ 
transitions. 
\item
It should also be stressed that these decays are of interest on 
its own as they are theoretically  cleaner than $B\to K^{(*)}\mu^+\mu^-$ and 
allow good tests of the presence of right-handed currents and in general of NP.
\end{itemize}

Both decays should be measured 
at Belle II. The most recent estimate of their branching ratios within the SM reads \cite{Buras:2014fpa}:
\be
\mathcal{B}(B^+\to K^+\nu\bar\nu) = \left[\frac{\vcb}{0.0409}\right]^2(3.98 \pm 0.43) \times 10^{-6}, 
\ee
\be
\mathcal{B}(B^0\to K^{* 0}\nu\bar\nu)  = \left[\frac{\vcb}{0.0409}\right]^2 (9.19\pm0.86) \times 10^{-6},
\ee
where the errors in the parentheses are fully dominated by form factor uncertainties. We expect that when these two branching ratios will be measured, these 
uncertainties will be further decreased and $\vcb$ will be precisely known so 
that a very good test of the SM will be possible.

An extensive analysis of these decays model independently and in various extensions of the SM has been performed in  \cite{Buras:2014fpa} but only the correlation of $\kpn$ with the $b\to s \nu\bar\nu$ in MFV can be found in figure~2 of that 
paper and we would like to extend this discussion.
 In view of the fact  that $B\to K^{(*)}\nu\bar\nu$ decays are correlated with  
$B\to K^{(*)}\mu^+\mu^-$ in $Z$ and $Z^\prime$ models and there are also correlations between 
 $B\to K^{(*)}\nu\bar\nu$ and $K\to\pi\nu\bar\nu$ decays in such models, we will find correlations between $\kpn$, $\klpn$ and $B\to K^{(*)}\mu^+\mu^-$ which can  be tested by LHCb and NA62 before Belle will test the correlations between  $B\to K^{(*)}\nu\bar\nu$ decays and 
$B\to K^{(*)}\mu^+\mu^-$ analyzed in detail  in  \cite{Buras:2014fpa}.

All formulae necessary for our analysis can be found in  \cite{Buras:2014fpa} 
and will not be repeated here (see in particular section 4.1 of that paper). 

\begin{figure}[t]
\centering%
\includegraphics[width=0.47\textwidth]{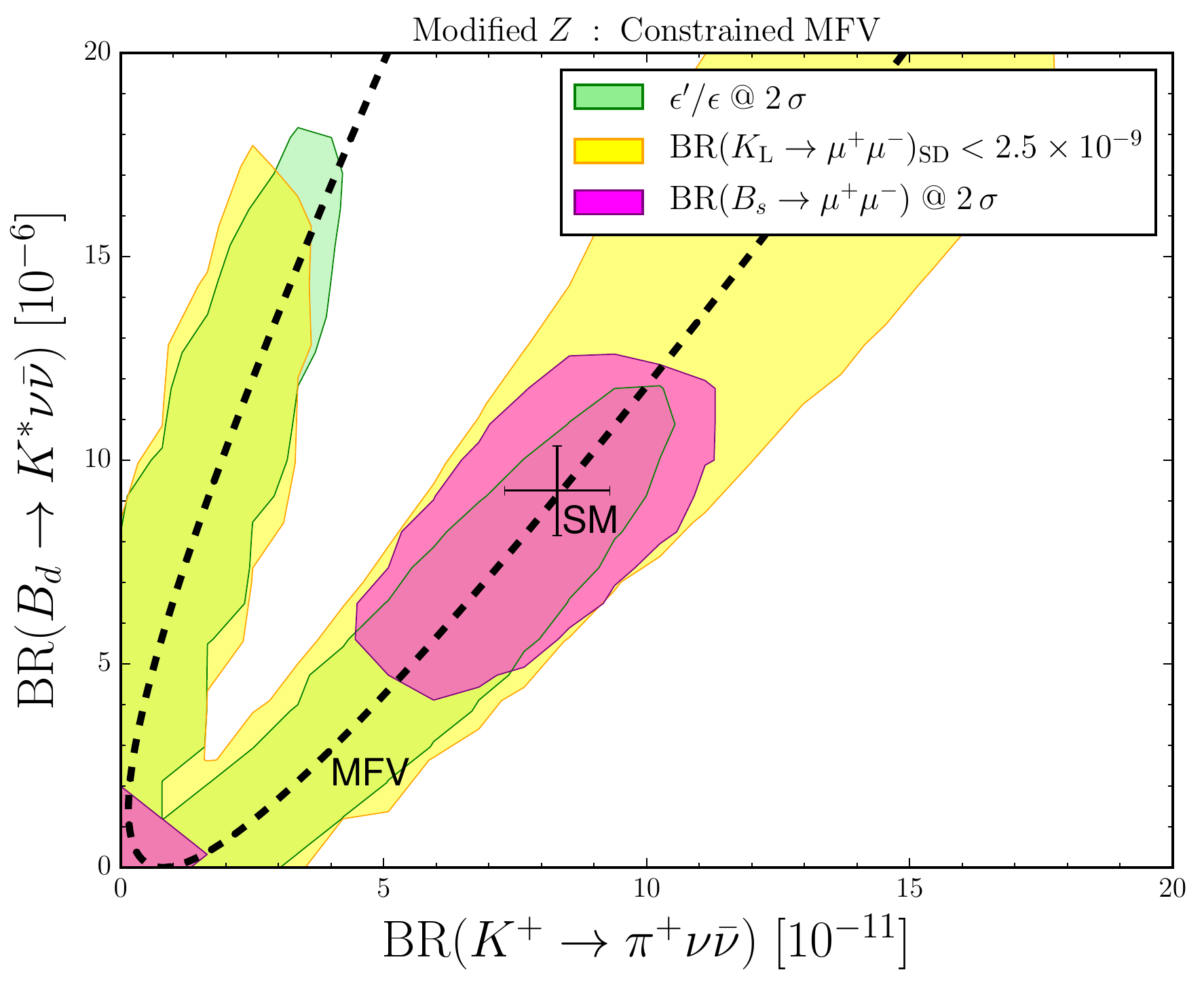}\hfill%
\includegraphics[width=0.47\textwidth]{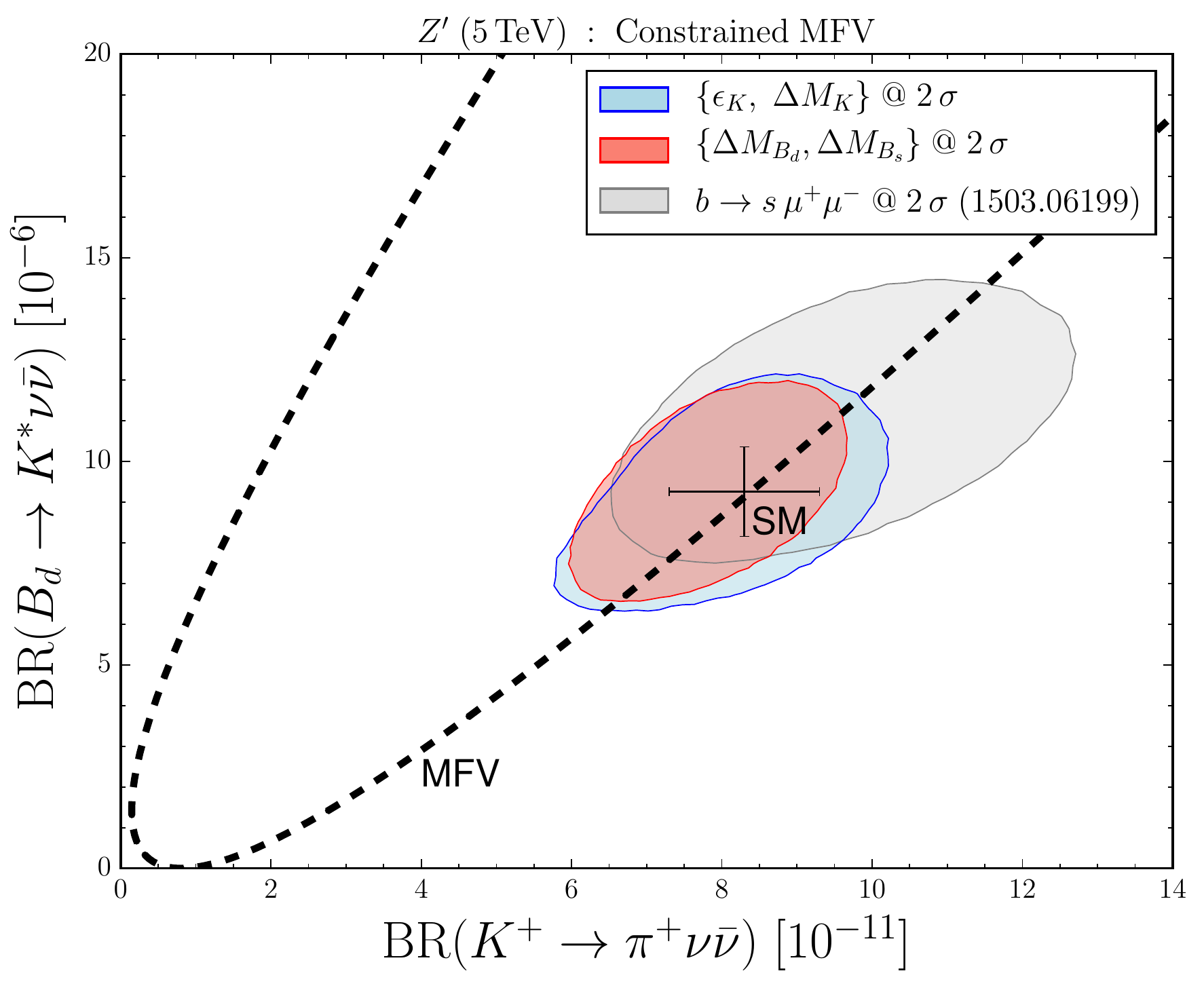}
\caption{\it Allowed ranges for $\mathcal{B}(\kpn)$ versus $\mathcal{B}(B_d\to K^*\nu\bar\nu)$ in a simplified $Z$ model (left panel) and a $5\tev~Z'$ model (right panel) obeying CMFV\@. In the left panel the $2\,\sigma$ confidence regions shown correspond to constraints from $\epsilon'/\epsilon$ (green), $K_{\rm L}\to \mu^+\mu^-$ (yellow) and $B_s \to \mu^+\mu^-$ (magenta), while in the right panel they correspond to constraints from kaon mixing (blue), $B$ mixing (red) and $b\to s\mu^+\mu^-$ transitions (grey) (from \cite{Altmannshofer:2015sma}).\label{fig:BKnn-CMFV}}
\end{figure}

In figure~\ref{fig:BKnn-CMFV} we show the regions allowed at 95\% C.L.\ in the $\mathcal{B}(\kpn)$ versus $\mathcal{B}(B_d\to K^*\nu\bar\nu)$ plane for a simplified $Z$ and a $5\tev~Z'$ model obeying CMFV. 
We do not show corresponding plots for $\mathcal{B}(B^+\to K^+\nu\bar\nu)$ because in CMFV the NP dependence is the same as for $\mathcal{B}(B_d\to K^*\nu\bar\nu)$.

\boldmath
\subsection{$K_L\to \mu^+\mu^-$}\label{sec:KLmm}
\unboldmath

Only the so-called short distance (SD)
part of a dispersive contribution
to $K_L\to\mu^+\mu^-$ can be reliably calculated. It is given 
generally as follows {($\lambda=0.2252$)}
\be
\mathcal{B}(K_L\to\mu^+\mu^-)_{\rm SD} = 2.01\cdot 10^{-9} 
\left( \frac{{\rm Re}\,Y_{\rm eff}}{\lambda^5} + \frac{{\rm Re}\,\lambda_c}{\lambda} P_c(Y)  \right)^2\,,
\ee
where at NNLO \cite{Gorbahn:2006bm}
\begin{align}
P_c(Y) &= 0.115\pm 0.017.
\end{align}
The short distance contributions are described by
\be\label{YK}
Y_{\rm eff} = V_{ts}^* V_{td} \left(Y_{L}(K) - Y_{R}(K)\right),
\ee
with
\begin{align}
Y_L^{\rm SM}(K) &= \eta_Y Y_0(x_t), & \eta_Y=0.9982,
\end{align}
also entering  $B_{s,d}\to\mu^+\mu^-$ decays. 
Notice the minus sign in front of $Y_R$, as opposed to $X_R$ in (\ref{XK}), that 
results from the fact that only the axial part contributes. 
This difference allows to be sensitive to right-handed couplings, which 
is not possible in the case of $K\to\pi\nu\bar\nu$ decays.

In the case of tree-level $Z$ exchange we have 
\begin{align}
Y_L(K)&=Y_L^{\rm SM}(K)+\frac{\Delta_A^{\mu\bar\mu}(Z)}{g^2_{\rm SM}M_{Z}^2}
                                       \frac{\Delta_L^{sd}(Z)}{V_{ts}^* V_{td}}, & 
Y_R(K)&=\frac{\Delta_A^{\mu\bar\mu}(Z)}{g^2_{\rm SM}M_{Z}^2}
                                       \frac{\Delta_R^{sd}(Z)}{V_{ts}^* V_{td}},
\end{align}
with analogous expressions for the $Z^\prime$ case.

If $Y(K)$ is related to $X(K)$, as in most of the models considered here, one can write $\B(K_L\to\mu^+\mu^-)$ in terms of $\B(\kpn)$ and $\B(\klpn)$, in analogy to \eqref{epsilonprimeKlpn}, as
\begin{align}
\B(K_L\to\mu^+\mu^-) &=  2.01\cdot 10^{-9} \left[\frac{{\rm Re}\,\lambda_t}{\lambda^5}(Y_{\rm SM} \mp X_{\rm SM}) + \frac{{\rm Re}\,\lambda_c}{\lambda}(P_c(Y) \mp P_c(X))\right.\notag\\
&\left.\quad\mp \left(\frac{\B(\kpn)}{\kappa_+} - \frac{\B(\klpn)}{\kappa_L}\right)^{1/2}\right],
\end{align}
where the first choice of signs holds whenever only left-handed contributions are present -- i.e. in MFV, $U(2)^3$, and in the LH scenario for generic couplings -- while the second choice holds for RH couplings. NP contributions to $\B(K_L\to\mu^+\mu^-)$ vanish in the LRS scenario for $Z$ and $Z'$.

The extraction of the short distance
part from the data is subject to considerable uncertainties. The most recent
estimate gives \cite{Isidori:2003ts}
\be\label{eq:KLmm-bound}
\mathcal{B}(K_L\to\mu^+\mu^-)_{\rm SD} \le 2.5 \cdot 10^{-9}\,,
\ee
to be compared with $(0.8\pm0.1)\cdot 10^{-9}$ in the SM.

As a preparation for the next section it is useful 
to recall what is the structure 
of the impact on $\kpn$ and $\klpn$ of the constraints from 
$\epe$ and $K_L\to\mu^+\mu^-$, which have an important interplay \cite{Buras:2014sba}:  $\epe$ puts 
constraints only on imaginary parts of NP contributions while $K_L\to\mu^+\mu^-$  only on the real ones. As demonstrated already  in 
\cite{Buras:2012jb}, the 
impact of the latter constraint on  $\kpn$ and $\klpn$ depends strongly 
on the scenario for the $Z$ flavour violating couplings.


\section{Results and comparison of bounds}\label{sec:6}

\subsection{Preliminaries}

The  detailed phenomenology  in the general case of $Z$ and $Z^\prime$ 
scenarios, including 
$\varepsilon_K$, $\Delta M_K$ and rare decays  $\kpn$, $\klpn$ and $K_L\to\mu^+\mu^-$, has been presented in \cite{Buras:2012jb} and  
generalized to include $\epe$ in \cite{Buras:2014sba}. But MFV 
has not been considered there and it will be of interest to see the allowed 
size of NP contributions in this case. Earlier studies of the upper bounds 
on NP effects in $\Delta F=2$ and $\Delta F=1$ processes can be found in 
\cite{Bobeth:2005ck,Haisch:2007ia}. Here we will concentrate on $\kpn$ and $\klpn$ decays but will also present some results for other decays. The analyses of rare processes in models with an $U(2)^3$ flavour symmetry has been already considered in \cite{Barbieri:2011fc,Barbieri:2012uh}, and in \cite{Buras:2012jb}  in the context of $Z$ and $Z'$ scenarios. But our analysis 
that uses simple models for couplings allows a new insight into these models.

Also, the present analysis uses a different strategy for the CKM parameters than 
the one in \cite{Buras:2014sba}, where various scenarios for these parameters 
have been considered. In what follows we will use the values of the parameters in (\ref{STRA}) determined 
in tree-level decays -- called ``strategy A'' in \cite{Buras:2015qea} -- and we will investigate how large NP effects in 
$\kpn$ and $\klpn$ are still allowed when the constraints from $\varepsilon_K$,  $\Delta M_K$,  $K_L\to\mu^+\mu^-$,  and $\epe$ are taken into account. As already described, the latter constraint will be 
subject to significant non-perturbative uncertainties connected to the parameter 
$\bsi$. In spite of this, $\epe$ already has an important impact on the maximal allowed size of the branching ratio, not only for $\klpn$ but also for $\kpn$.

In fact the recent progress on the calculation of $\epe$ in \cite{Buras:2015yba} and \cite{Buras:2015xba}, reported already in section~\ref{sec:3a}, makes the impact of this ratio on rare decays larger than in \cite{Buras:2014sba}.
In the following we shall use the lattice value $B_8^{(3/2)} = 0.76(5)$ from \cite{Blum:2015ywa}, while for $B_6^{(1/2)}$ we will take an average between the new lattice result \cite{Bai:2015nea} and the maximal value $B_6^{(1/2)} = B_8^{(3/2)}$ allowed by the large $N$ approach \cite{Buras:2015xba}.

In table~\ref{tab:input} we summarise the values of the parameters used as inputs in our analysis.

\begin{table}[!tb]
\renewcommand{\arraystretch}{1.2}
\centering%
\begin{tabular}{|cl|cl|}
\hline\hline
$|V_{ub}|$ & $3.88(29)\times 10^{-3}$\hfill\cite{Buras:2015qea}
&
$F_K $ & $ 156.1(11)\mev$\hfill\cite{Aoki:2013ldr}
\\
$|V_{cb}|$ & $40.7(14)\times 10^{-3}$\hfill\cite{Buras:2015qea}
&
$\hat{B}_K$ & $0.750(15)$ \hfill \cite{Aoki:2013ldr,Buras:2014maa}
\\
$\gamma$ & $\left(73.2^{+6.3}_{-7.0}\right)^\circ $\hfill\cite{Trabelsi:2014}
&
$F_{B_d}$ & $190.5(42)\mev$ \hfill \cite{Aoki:2013ldr} 
\\
$|V_{us}|$ & $0.2252(9)$\hfill\cite{Amhis:2012bh} 
& 
$F_{B_s}$ & $227.7(45)\mev$ \hfill \cite{Aoki:2013ldr} 
\\\cline{1-2}
$|\eps_K|$ & $2.228(11)\times 10^{-3}$\hfill\cite{Beringer:1900zz} 
& 
$F_{B_s}\sqrt{\hat{B}_{B_s}}$ & $266(18)\mev$\hfill\cite{Aoki:2013ldr}
\\
$\Delta M_K$ & $ 0.5292(9)\times 10^{-2} \,\text{ps}^{-1}$\hfill\cite{Beringer:1900zz} 
& 
$\xi $ & $1.268(63)$\hfill\cite{Aoki:2013ldr}
\\
$\Delta M_d $ & $ 0.507(4)\,\text{ps}^{-1}$\hfill\cite{Amhis:2012bh} 	  
& 
$B_6^{(1/2)}$& 0.65(20)\hfill\cite{Bai:2015nea,Buras:2015xba}
\\
$\Delta M_s $ & $ 17.761(22)\,\text{ps}^{-1}$\hfill\cite{Amhis:2012bh}
&
$B_8^{(3/2)}$& 0.76(5)\hfill\cite{Blum:2015ywa}
\\
\cline{3-4}
$\tau_{B_d}$ & $ 1.519(5) \,\text{ps}$\hfill\cite{Amhis:2012bh} 
&
$\eta_{cc}$ & $1.87(76)$\hfill\cite{Brod:2011ty}
\\
$\tau_{B_s}$ & $ 1.512(7)\,{\rm ps}$\hfill\cite{Amhis:2012bh}
&
$\eta_{ct}$ & $ 0.496(47)$\hfill\cite{Brod:2010mj}
\\
\cline{1-2}
$\alpha_s(M_Z)$ & $0.1185(6) $\hfill\cite{Beringer:1900zz}
&
$\eta_{tt}$ & $0.5765(65)$\hfill\cite{Buras:1990fn}
\\
$m_c(m_c) $ & $ 1.279(13) \gev$ \hfill\cite{Chetyrkin:2009fv}
&
$\eta_B$ & $0.55(1)$\hfill\cite{Buras:1990fn,Urban:1997gw}\\
$m_s(2\gev)$ & $93.8(24)\mev$\hfill\cite{Aoki:2013ldr}& & \\
$m_d(2\gev)$ & $4.68(16)\mev$\hfill\cite{Aoki:2013ldr}& & \\
$M_t $ & $ 173.34(82)\gev$\hfill\cite{ATLAS:2014wva}& & \\
\hline\hline
\end{tabular} 
\caption{\it Values of theoretical and experimental quantities used as input parameters.}\label{tab:input}
\end{table}


\boldmath
\subsection{CMFV and $U(2)^3$ for $Z$  and $Z'$ models}
\unboldmath

\begin{figure}[t]
\centering%
\includegraphics[width=0.47\textwidth]{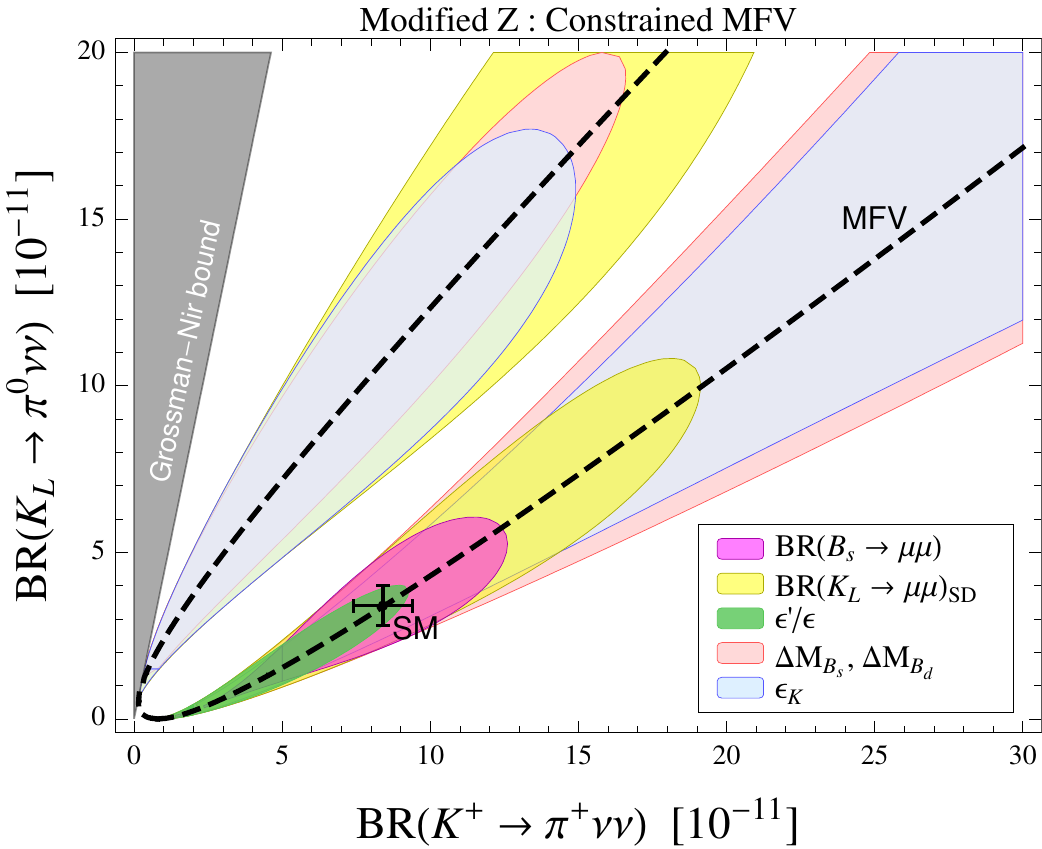}\hfill%
\includegraphics[width=0.47\textwidth]{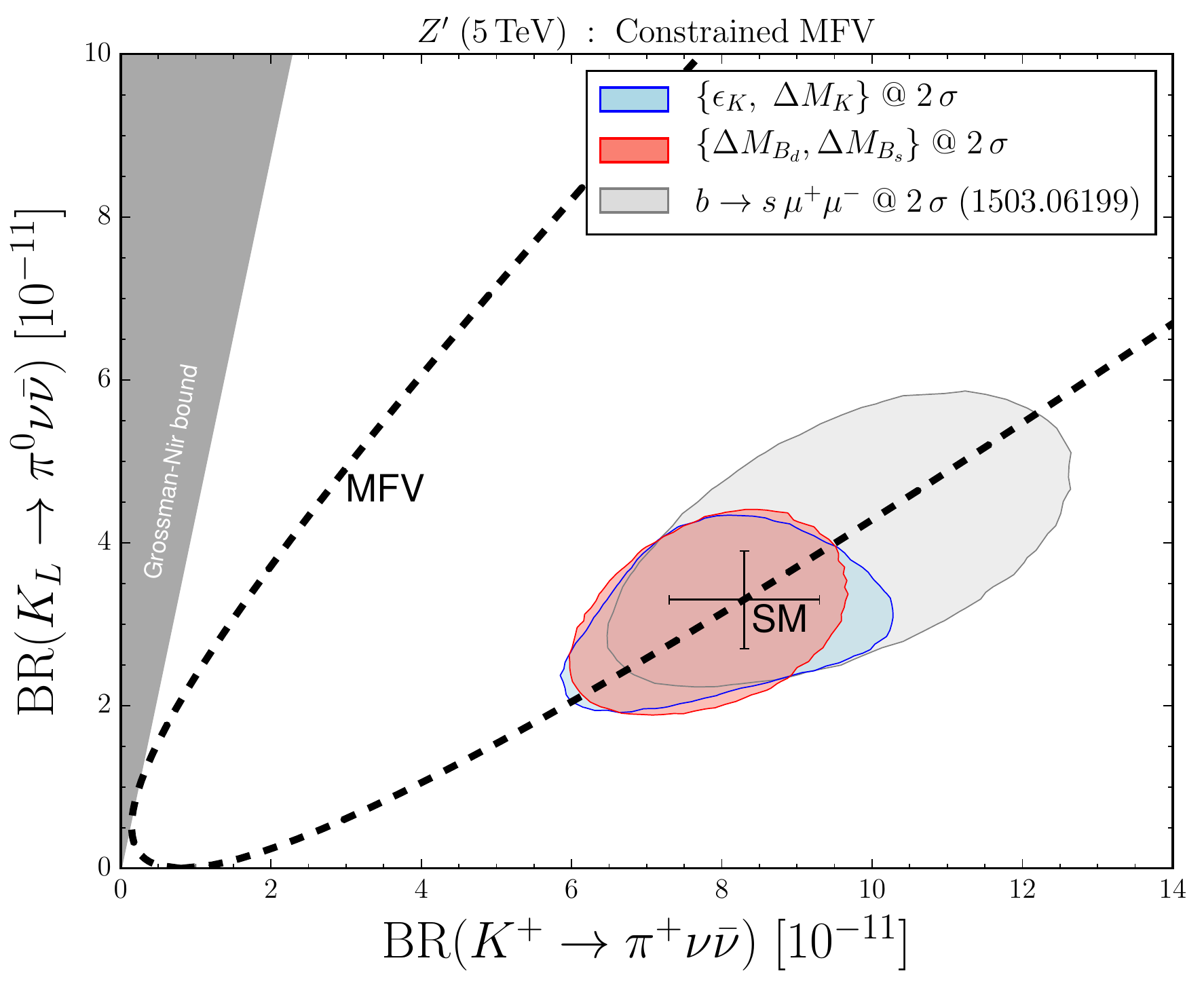}
\caption{\it The 95\% C.L. allowed ranges for  $\mathcal{B}(\klpn)$ and $\mathcal{B}(\kpn)$ in a simplified $Z$ model (left panel) or a $5\tev~Z'$ model (right panel) obeying CMFV\@. In the case of the smaller $U(2)^3$ symmetry, the constraints from $B$ processes can be neglected. Note the difference in scale between these plots.\label{fig:Knn-CMFV}}
\end{figure}

In the left panel of figure~\ref{fig:Knn-CMFV} we show the $2\,\sigma$ allowed ranges from current experimental constraints for $\mathcal{B}(\klpn)$ and $\mathcal{B}(\kpn)$ in a simplified $Z$ model obeying CMFV\@.  
Similarly, in the right panel of the same figure, we show the allowed ranges for a simplified $Z'$ model with a $Z'$ mass of $5\tev$, as discussed earlier, also obeying CMFV.
Neglecting the constraints from the $B_{d,s}$ systems gives the situation in the less constrained $U(2)^3$ symmetry scenario.
In both cases we have used the averaged CKM inputs from strategy A\@.
We make the following observations:
\begin{itemize}
    \item For the simplified $Z$ model, constraints from $\Delta F=1$ processes dominate over $\Delta F=2$ ones. The latter in fact hardly constrain these branching ratios at all.
    \item For $Z'$ models the situation is the opposite: due to a direct dependence on the high NP scale, $\Delta F=2$ observables become the most constraining, and we have therefore neglected the $\Delta F=1$ constraints. 
    \item NP contributions in simplified $Z$ models with CMFV are rather constrained by the $B_s\to \mu^+\mu^-$ branching ratio. 
In $U(2)^3$ this constraint is not present, while the short distance part of $K_{\rm L}\to \mu^+\mu^-$ still leaves ample room for NP.
On the other hand, the strongest limit for an enhancement of $\B(\kpn)$ and 
$\B(\klpn)$ branching ratios, both in $U(3)^3$ and $U(2)^3$ Z models, comes from $\epe$. Indeed, already the SM point is only marginally compatible with the experimental data, and lower values of the two branching ratios are preferred.
    \item For $Z'$ models the $\Delta F=2$ constraints from the kaon and $B$ systems are comparable in size, therefore there is little difference between the CMFV and $U(2)^3$ scenarios.
For a $5\tev$ $Z'$ they can deviate from the SM  by at most $10-20\%$, which 
could be hard to detect even in the flavour precision era.
\end{itemize}

In summary we find that it will not be easy to distinguish MFV models from the SM on the basis of $\kpn$ and $\klpn$. While in the case 
of $Z^\prime$ models small NP effects are required by $\Delta F=2$ constraints,  because of the high $Z^\prime$ mass, in the case of $Z$ models the crucial limit comes from the data on $B_s\to \mu^+\mu^-$ and $\epe$. While an enhancement of the two branching ratios is always strongly constrained, their suppression with respect to the SM prediction is still possible in the latter case.


\boldmath
\subsection{Generic $Z$ models}
\unboldmath

\begin{figure}[t]
\centering%
\includegraphics[width=0.47\textwidth]{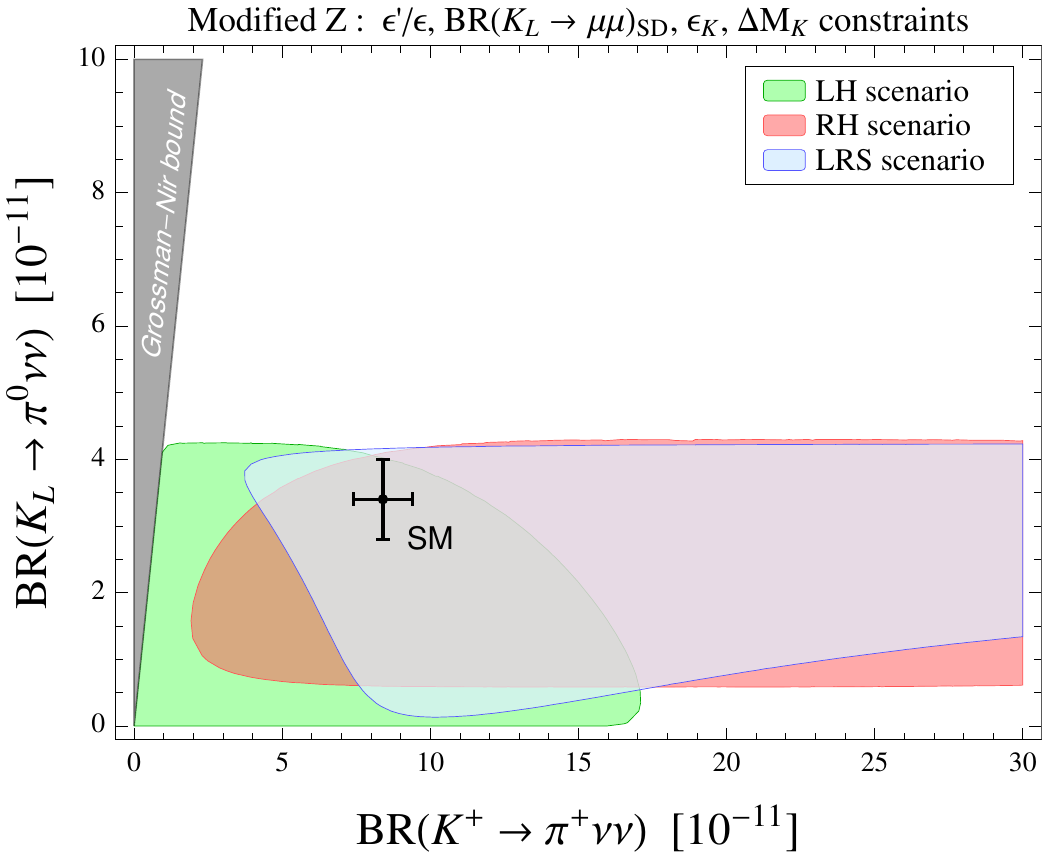}\hfill%
\includegraphics[width=0.47\textwidth]{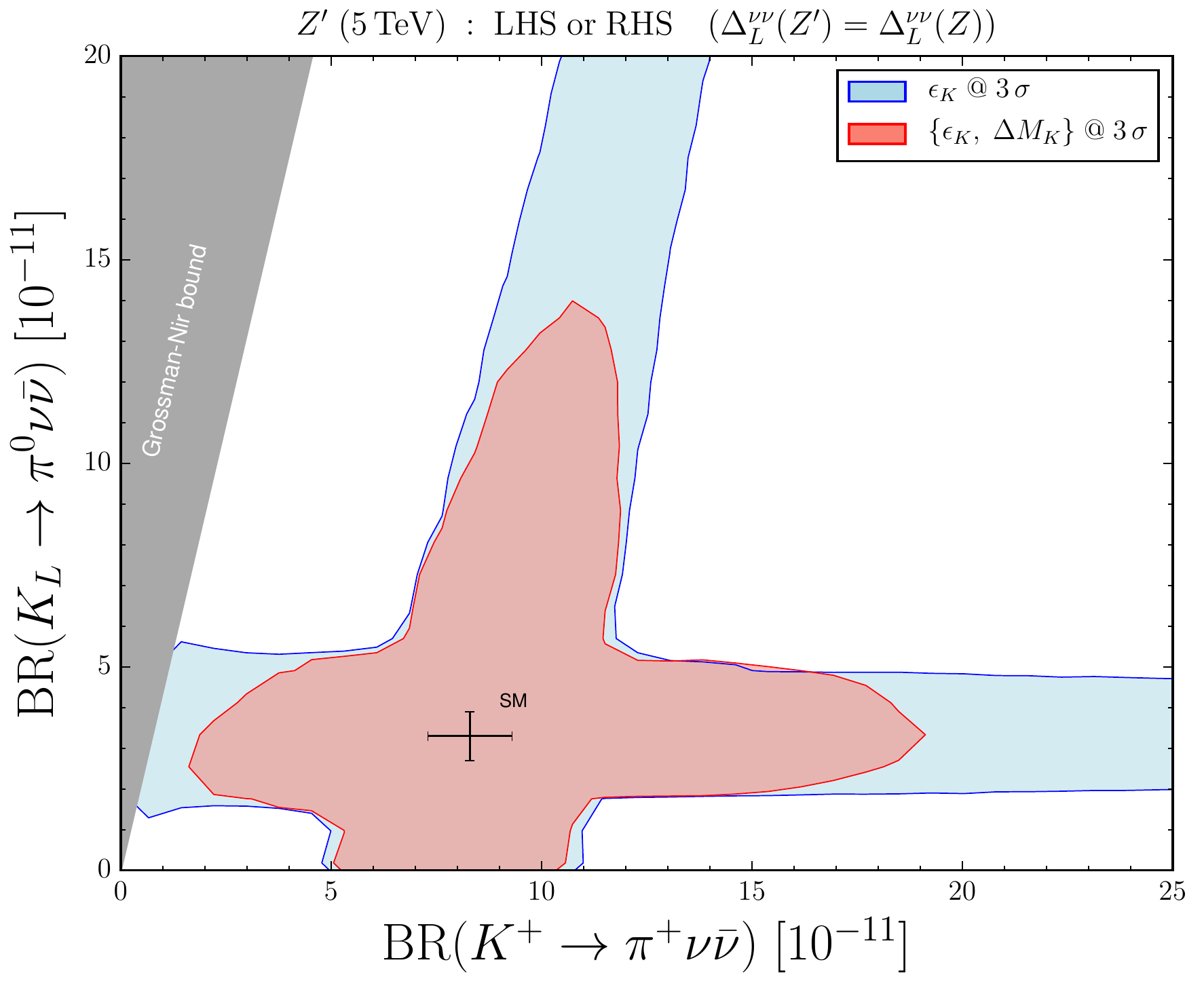}
\caption{\it The allowed ranges for  $\mathcal{B}(\klpn)$ and $\mathcal{B}(\kpn)$ in a simplified $Z$ model (left) and a $5\tev~Z'$ model (right) in LH and RH scenarios. The $\varepsilon_K$  and $\Delta M_K$ constraints are imposed in all cases. In the left-handed plot the $\epsilon'/\epsilon$ and $K_L\to\mu\mu$ constraints are also imposed. \label{fig:Knn_Zp-LHS}}
\end{figure}

In the left panel of figure~\ref{fig:Knn_Zp-LHS} we show the 95\% C.L.\ allowed ranges for  $\mathcal{B}(\klpn)$ and $\mathcal{B}(\kpn)$ in the LH, RH and LR  scenarios with Z mediated FCNC. The origin 
for the different ranges is explained in detail in  \cite{Buras:2014sba}. 
Here we only note the following basic features:

\begin{itemize}
\item
In the LH scenario $\mathcal{B}(\kpn)$ can be by a factor of two larger than 
its SM value. The strong $\epe$ constraint, on the other hand, forces $\mathcal{B}(\klpn)$ to be of the order of the SM value or smaller, as explained in section~\ref{sec:lat}. Both branching ratios can also be significantly suppressed. We show the impact of 
the $\epe$ and $K_L\to \mu^+\mu^-$ constraints. 
\item
In the RH scenario $\mathcal{B}(\klpn)$ is again constrained to be close to
its SM value, while $\mathcal{B}(\kpn)$ can be almost by a factor of five larger than its 
SM value because the $K_L\to\mu^+\mu^-$ constraint is weaker. Such a large enhancement is anyhow already constrained by the present experimental results. Both 
branching ratios can also be suppressed relative to SM values but not as 
strongly as in the LHS case. 
\item
Finally in the LRS case the allowed range for $\mathcal{B}(\klpn)$ is similar 
to the RHS case, while, due to the absence of the $K_L\to\mu^+\mu^-$ constraint, $\mathcal{B}(\kpn)$ can be large. The $\epsilon_K$ constraint plays a role here because of the presence of left-right operators.
\end{itemize}


\boldmath
\subsection{Generic $Z^\prime$ models}
\unboldmath

Due to the sensitivity of the $\epe$ constraint to $Z^\prime$ diagonal quark 
couplings, in order to be model independent, we present a numerical 
analysis in $Z^\prime$ scenarios without the $\epe$ constraint.
In the right panel of figure~\ref{fig:Knn_Zp-LHS} we show the $3\,\sigma$ allowed ranges for  $\mathcal{B}(\klpn)$ and $\mathcal{B}(\kpn)$ in a simplified $5\tev~Z'$ model for the LH scenario obeying the $\varepsilon_K$ and $\Delta M_K$ constraints. The leptonic $Z'$ couplings have been fixed to the $Z$ boson values for concreteness, $\Delta_L^{\nu\bar\nu}(Z') = \Delta_L^{\nu\bar\nu}(Z)$.
Since the $\Delta F = 2$ effects due to RH currents alone are identical to the ones of LH currents, exactly the same results hold also in the RH scenario.

In the LRS scenario the constraints from $\epsilon_K$ are much stronger, due to the presence of left-right operators. Notice, on the other hand, that one can in principle avoid the strong $\Delta F = 2$ bounds by means of some fine-tuning if the RH couplings are sufficiently small \cite{Buras:2014zga}; we do not analyse this possibility here.


\section{Summary and Outlook}\label{sec:8} 

In the present paper we have made another look at $\kpn$ and $\klpn$ decays 
which are expected to become the stars of flavour physics in the coming ten 
years. Our results are presented in numerous plots which should allow to monitor  efficiently the experimental developments in the coming years. In particular 
the correlations with other observables like $B_{s,d}\to\mu^+\mu^-$, $B\to K(K^*)\mu\bar\mu$  and 
$B\to K(K^*)\nu\bar\nu$ branching ratios and $\epe$ will be very relevant 
for the distinction between various extensions of the SM. Also the improvement 
in the accuracy of the CKM parameters determined in tree-level decays and more
 accurate values of various non-perturbative parameters obtainted by lattice 
QCD will be important ingredients in future analyses.

In view of the recent result on $\epe$ from RBC-UKQCD collaboration \cite{Bai:2015nea} and the 
analyses in \cite{Buras:2015yba,Buras:2015xba} which 
find $\epe$ significantly below the data, we have presented two 
simplified models 
which would improve the agreement of the theory and data if the present status of $\epe$ will be confirmed by more precise lattice QCD calculations one day.

We close our paper with the following observations:
\begin{itemize}
\item
There is a hierarchy in the size of possible NP effects in $K\to\pi\nu\bar\nu$ 
mediated by tree-level $Z$ exchanges. 
They are smallest in CMFV, larger in $U(2)^3$ models and significantly larger 
in the case of new sources of flavour and CP violation beyond these two CKM-like
frameworks.
\item
In $Z^\prime$ models with MFV the present $B_d\to K (K^*) \mu^+\mu^-$ anomalies favour the enhancement of $\kpn$ and $\klpn$.
$\Delta F = 2$ observables however put significant constraints on this possibility.
\item
Due to the absence of correlation between $K\to\pi\nu\bar\nu$ and $\epe$ 
in general $Z^\prime$ models, the size of NP contribution in these decays could be large. Then, as demonstrated
in \cite{Buras:2014zga}, $\kpn$ and $\klpn$ can probe energy scales as large as $1000\tev$ in the presence of general 
flavour-violating couplings.
\item
If the NA62 experiment will find the branching ratio for $\kpn$ to be 
significantly above the SM predictions, both 
tree-level $Z$ and $Z^\prime$ exchanges could be responsible for these effects -- but the same can be said about more complicated models like LHT, RSc and 
supersymmetric models. Such high values of $\mathcal{B}(\kpn)$ will also signal non-MFV sources at work. 
\item
In particular, only $Z$ and $Z^\prime$ models with general flavour violating couplings, among the models that we considered, allow 
for $\mathcal{B}(\kpn)$ above
$20\times 10^{-11}$.
\item
Finally, the future measurement of $\mathcal{B}(\klpn)$ will significantly 
facilitate the distinction between various models.
\end{itemize}


\section*{Acknowledgements}

We would like to thank Jennifer Girrbach-Noe for participation in very early 
stage of this project.
This research was done and financed in the context of the ERC Advanced Grant project ``FLAVOUR''(267104) and was partially
supported by the DFG cluster
of excellence ``Origin and Structure of the Universe''.

\bibliographystyle{JHEP}
\bibliography{allrefs}
\end{document}